\documentclass[review,3p,sort&compress]{elsarticle}

\usepackage{amsmath,amssymb,amsfonts}
\usepackage{graphicx}
\usepackage{textcomp}
\usepackage{xcolor}

\usepackage[hidelinks]{hyperref}
\usepackage[utf8]{inputenc}
\usepackage{booktabs} 
\usepackage{microtype}
\usepackage{tabularx} 
\usepackage{array}
\usepackage{paralist}
\usepackage{eurosym}
\usepackage{pbox} 
\usepackage{multirow}
\usepackage{subcaption}
\usepackage{stackengine}
\usepackage{adjustbox}
\usepackage[noend]{algpseudocode}
\usepackage{algorithm}
\usepackage{algorithmicx}
\usepackage{setspace}
\usepackage{xspace}
\usepackage{siunitx}

\newtheorem{definition}{Definition}

\newcommand{\mypar}[1]{\smallskip\noindent\textbf{#1.}}
\newcommand{\mypartwo}[1]{\smallskip\noindent\textit{#1.}}

\newcommand{\minedcoll}{C_{\mathcal{M}}}
\newcommand{\minedconst}{c_{\mathcal{M}}}

\newcommand{\instcoll}{C_{L}}
\newcommand{\instconst}{c_{L}}

\newcommand{\papertitle}{Mining Constraints from Reference Process Models \\for Detecting Best-Practice Violations in Event Logs}

\journal{Information Systems}
\begin{document}

\begin{frontmatter}

\title{\papertitle}

\author[uma,sap]{Adrian Rebmann\corref{mycorrespondingauthor}}
\cortext[mycorrespondingauthor]{Corresponding author}
\ead{rebmann@uni-mannheim.de}
\author[sap]{Timotheus Kampik}
\ead{timotheus.kampik@sap.com}
\author[uko]{Carl Corea}
\ead{ccorea@uni-koblenz.de}
\author[uma]{Han van der Aa}
\ead{han.van.der.aa@uni-mannheim.de}

\address[uma]{University of Mannheim, B6 26, 68159 Mannheim, Germany}
\address[sap]{SAP Signavio, Berlin, Germany}
\address[uko]{University of Koblenz, Koblenz, Germany}

\date{Received: September 4, 2023 }

\begin{abstract}
    Detecting undesired process behavior is one of the main tasks of process mining and various conformance-checking techniques have been developed to this end. These techniques typically require a normative process model as input, specifically designed for the processes to be analyzed. Such models are rarely available, though, and their creation involves considerable manual effort.
    However, reference process models serve as best-practice templates for organizational processes in a plethora of domains, containing valuable knowledge about general behavioral relations in well-engineered processes. These general models can thus mitigate the need for dedicated models by providing a basis to check for undesired behavior. 
    Still, finding a perfectly matching reference model for a real-life event log is unrealistic because organizational needs can vary, despite similarities in process execution.
    Furthermore, event logs may encompass behavior related to different reference models, making traditional conformance checking impractical as it requires aligning process executions to individual models.
    To still use reference models for conformance checking, we propose a framework for mining  declarative best-practice constraints from a reference model collection, automatically selecting constraints that are relevant for a given event log, and checking for best-practice violations.
    We demonstrate the capability of our framework to detect best-practice violations through an evaluation based on real-world process model collections and event logs.
\end{abstract}


\end{frontmatter}

\section{Introduction}
\label{sec:introduction}
\emph{Process mining} is a discipline that analyzes event data recorded during the execution of organizational processes, with the aim of gaining insights into how these processes are actually running~\cite{van2022process}. 
One of the main tasks of process mining is the detection of \emph{undesired} process behavior, as it can reveal compliance problems, process inefficiencies, and data quality issues.
Mainly, such undesired behavior can be detected using \emph{conformance-checking} techniques~\cite{carmona2018conformance}, which require a normative process model that was specifically created for the process to be analyzed. Here, the normative process model is compared with event data in order to reveal any deviations that might have occurred. While such approaches are very valuable for companies, an inherent problem is that such normative process models are rarely available in organizations and require time-consuming and costly efforts in their creation~\cite{dumas2018fundamentals}. 
Fortunately, many process types, such as procurement and invoicing processes, are commonly organized in similar ways across organizations or---at the very least---involve similar steps. 
As a consequence, \emph{reference process models} have been recognized as an important means to provide depictions of proven ways to run these processes, serving as best-practice templates for process implementations in various domains ~\cite{fettke2003classification, gottschalk2008mining}. Beyond the best practices such models capture for entire process types, they also contain important knowledge about general behavioral relations that have to hold in these processes, e.g., that an invoice must be checked before it is approved.
The availability of such reference process models can, thus, alleviate the need for manually creating a customized normative model, as the former can serve as a basis against which to check for undesired behavior in the scope of conformance checking.

A core problem, however, is that it is generally unrealistic to find a single reference process model that exactly fits to a given real-life process, because, even though the general way of executing such processes may be similar, the individual needs of organizations can vary~\cite{gottschalk2008mining}. This means that a real-world process may be subject to additional requirements than the ones captured in a reference model; also, some parts of a reference model may not be applicable in a particular situation.
Additionally, an event log might capture behavior that relates to multiple reference models, e.g., related to both procurement and invoicing.
 This makes traditional conformance checking involving a singular normative model hard for companies. Therefore, in order to leverage the best practices captured in the reference models for the detection of undesired behavior, the alignment between these and event data needs to be achieved in a more flexible manner, involving multiple reference models at once. 
 Nevertheless, it also brings additional challenges, e.g., identifying which reference models are relevant.

We approach this problem through the development of a framework for the extraction of so-called \emph{relevant} declarative constraints~\cite{DiCiccio2022} from a collection of reference models. Once extracted, they can be used to check if event logs violate best practices captured in the constraints. The constraints we extract may stem from thousands of reference process models that capture best-practices of a plethora of domains, depending on the model collection used as input.
Here, the main challenge is that we need to determine which constraints are actually relevant and interesting---or even applicable---for a given event log. 
We address this challenge through established techniques from declarative process mining, as well as natural-language-processing techniques to refine and measure the relevance of mined constraints for a given event log. In this way, our framework can take advantage of repositories of reference models and identify relevant best-practice constraints that can be used for analyzing a given event log.
We implement our framework and demonstrate its capability to detect best-practice violations through an evaluation based on real-world process model collections.

The remainder of this paper is structured as follows. \autoref{sec:motivation} motivates the goal of mining constraints from reference process models to check for best-practice violations and \autoref{sec:preliminaries} defines preliminaries.
\autoref{sec:framework} presents our framework for mining, selecting, and checking best-practice constraints, which we evaluate in \autoref{sec:evaluation}. \autoref{sec:reallife} presents application scenarios, where we apply our framework to real-life event data.
Finally, \autoref{sec:related} discusses related work and \autoref{sec:conclusion} concludes the paper.

\section{Motivation}
\label{sec:motivation}
As a motivational example, this section illustrates the potential of checking for best-practice violations using declarative constraints that are mined from reference process models. 

\begin{figure}[!h]

    \centering
    \includegraphics[width=0.8\textwidth]{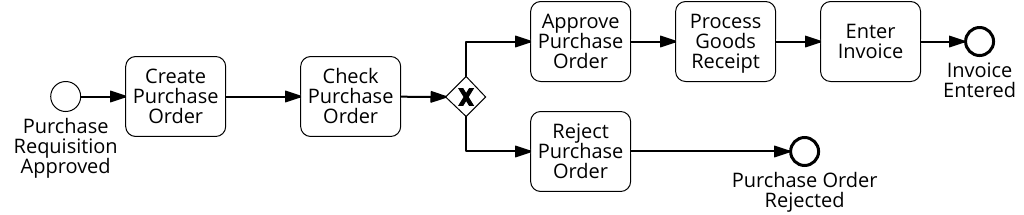}
    \caption{Exemplary reference process model in BPMN.}
    \label{fig:examplemodel}
\end{figure}


\mypar{Recognizing best-practice violations using a reference process model}
Consider the process model in \autoref{fig:examplemodel}, which shows an exemplary reference process model of a procurement process for materials. 
From this model, we can extract a variety of behavioral relations, such as:
A \textit{purchase order (PO)} must be \textit{created} before it is \textit{checked}
and, also, \textit{checked} before it is \textit{approved}; a \textit{PO} must not be both \textit{approved} and \textit{rejected}; a \textit{PO} must be \textit{approved} before \textit{goods} are \textit{received}; and no \textit{invoice} should be \textit{entered} when a \textit{PO} is \textit{rejected}.

\noindent
Now, consider a trace\footnote{Intuitively, a trace is a sequence of events that represents a process execution. We will formalize this in Section \ref{sec:preliminaries}.} $\sigma$ that stems from an event log of a procure-to-pay process for consumer products:

\smallskip
$ \sigma = \langle$\textit{create order, enter goods receipt, enter invoice, check invoice amount, make payment}$\rangle$

\smallskip

\noindent
The trace can be understood such that first an \textit{order} is \textit{created}, then a \textit{goods receipt} is \textit{entered}, and so forth. We can use the relations extracted from the reference process model to identify behaviors in $\sigma$ that violate best practices.
For instance, \textit{goods} are \textit{received} in $\sigma$ but an \textit{order }was neither \textit{checked} nor \textit{approved} beforehand and an invoice is \textit{entered} before \textit{goods} were \textit{received}. While the latter may depend on the specific organizational context, e.g., an arrangement with a supplier would allow for entering an invoice before goods are delivered, the former is clearly undesirable, regardless of any specific arrangements. 

The example thus shows that the reference model can be used to gain insights on best-practice violations of the given trace, despite not stemming from exactly the same process.
In fact, some of the behavioral relations captured in reference models can be broadly applicable. For example, the fact something cannot be both \textit{approved} and \textit{rejected} applies to virtually any process in which such approval decisions need to be made.

\mypar{Combining constraints from multiple reference process models} An important thing to recognize, though, is that, besides containing activities that relate to materials procurement,  trace $\sigma$ also contains activities that go beyond the scope of the reference process model shown in \autoref{fig:examplemodel}. In particular, $\sigma$ also contains activities related to \emph{checking the invoice amount} and \emph{making a payment}, which would sooner be covered by a reference model relating to an invoicing process.
Therefore, relying solely on a single reference model may result in missing relevant best-practice violations, which means that it is necessary to jointly check for adherence to best practices captured in multiple reference models.





This can be achieved by turning the behavioral relations captured by imperative process models (such as \autoref{fig:examplemodel}) into declarative constraints. This has two main benefits: first, it allows one to consider behavioral relations from multiple reference models at the same time, while, second, it allows one to omit parts of the imperative process model, that are not relevant to the trace at hand, from consideration. 
Specifically, in the presence of tens of thousands of constraints, we need to avoid that constraints that are irrelevant for an event log cause the detection of violations, i.e., \emph{false positives}.
For instance, a constraint such as ``\emph{process goods receipt} should precede \emph{enter invoice}'' (extracted from a procurement process model) should not be applied to a sales process event log, which does not involve \textit{goods receipt} objects.



However, expecting users to manually select from tens of thousands of constraints is unfeasible. To reduce manual effort, we need to automatically preselect constraints based on the specific context, i.e., the given event log. 
To this end, our framework mines constraints from a large collection of a reference process models and, subsequently, selects those constraints that are actually relevant to the particular situation at hand. 


\section{Preliminaries}
\label{sec:preliminaries}
We continue with preliminaries on event data, process models, and declarative constraints.
 
\subsection{Traces, Events, Event Logs}
Let $\mathcal{A}$ refer to the universe of possible activities in organizational processes and $\mathcal{R}$ to the universe of organizational roles that can be involved in their execution. Then, a trace $\sigma$ is a sequence that represents the execution of an organizational process. Such a trace consists of a finite sequence of events $\sigma = \langle e_1,...,e_n\rangle$, where each event $e_i$ is a tuple $(a, r)$, with $a \in \mathcal{A}$ as the activity that the event relates to and $r \in \mathcal{R} \cup \{\bot\}$ as a possibly empty attribute that indicates the role of the resource that executed this activity.
Note that, in the remainder, we commonly use dot notation to refer to components of tuples, e.g., using $e_i.a$ as a shorthand to refer to the activity of an event $e_i$

An event log $L$ is then a finite multi-set of traces.
 Given a trace $\sigma$, we use $\sigma^a$ to refer to the sequence of its activities, i.e., $\sigma^a = \langle e_1.a, \dots, e_n.a \rangle$. 
$L^a$ is then the multi-set of activity sequences obtained from all traces in $L$, i.e., $L^a = \{\sigma^a \mid \sigma \in L\}$. Finally, we use $A_L \subseteq \mathcal{A}$, to denote the set of activities that the events of the traces in $L$ relate to, and $R_L$ to denote the set of resource roles executing activities in $L$.

\subsection{Process Models and Process Model Collections}
Using $\mathcal{M}$ to denote a process model collection, each process model $M \in \mathcal{M}$ is a tuple $M=$($A$, $F$, $R$, $D$). $A$ is the set of labeled steps in $M$, including tasks and events in a BPMN diagram or transitions in a Petri net. Just like for events, we call $a \in A \subseteq \mathcal{A}$ an \emph{activity} in the remainder. 
$F$ is the set of unique finite execution sequences allowed by the model, with each $\pi \in F = \langle a_1,...,a_n\rangle$ as a sequence of activities in $A$.
Focusing on finite traces captures the intuition that each process execution is expected to complete in a finite number of steps~\cite{DiCiccio2022}.
$R \subseteq \mathcal{R}$ is a set of roles involved in the process represented by $M$ and, finally, $D: \mathcal{A} \rightarrow \mathcal{R} \cup \{\bot\}$ is a mapping that associates activities with roles, e.g., a \emph{create invoice} activity may be mapped to a \emph{vendor} role in a procurement process model.

\subsection{Declarative Constraints}
To specify best-practice constraints, we adopt concepts of \textsc{Declare} and \textsc{MP-Declare}.
\textsc{Declare} is a modeling language for declarative constraints~\cite{pesic2007declare} that describes a set of templates that restrict occurrences and possible orderings of activities in a process. 
\textsc{MP-Declare} is an extension of \textsc{Declare}~\cite{burattin2016conformance} and can express constraints over multiple perspectives of a process, e.g., the resource and data perspectives, using \textsc{Declare} templates. The semantics of \textsc{Declare} templates are often specificied with Linear Temporal Logic (LTL), which allows to exploit the amenities of LTL verification, with the full complexity of LTL ``hidden" from the user. \autoref{table:declare} summarizes commonly used templates and their LTL semantics~\cite{DiCiccio2022}. 

\begin{table}[!h]
    \centering
    \small 
    \caption{Semantics of \textsc{Declare} templates}
    \label{table:declare}
\begin{tabular}{r p{5cm} c}
  \toprule
  \textbf{Template} & \textbf{LTL semantics} & \textbf{Activation} \\
  \midrule
  \textsc{AtLeastOne}(a) & \textbf{F} a & at start \\
  \textsc{AtMostOne}(a) & $\neg\textbf{F}(a \land \textbf{X}(\textbf{F} a))$ & at start \\
  \textsc{ExactlyOne}(a) &  \textsc{AtLeastOne}(a) $\land$ \textsc{AtMostOne}(a) & at start \\
  \textsc{Absence}(a) & $\neg$\textbf{F} a & at start \\
  \midrule
  \textsc{RespondedExistence}(a,b) & \textbf{F} a $\rightarrow$ \textbf{F} b & a \\
  \textsc{Response}(a,b) & \textbf{G}(a $\rightarrow$ \textbf{F} b) & a \\
  \textsc{AlternateResponse}(a,b) & $ \textbf{G}(a \rightarrow$ $\textbf{X}(\neg a \textbf{U} b))$ & a \\
  \midrule
  \textsc{Precedence}(a,b) &\textbf{G}(b $\rightarrow$ \textbf{O} a) & b \\
  \textsc{AlternatePrecedence}(a,b) &$(\neg b$ \textbf{S} $a) \land $\textbf{G} $(b \rightarrow (\neg b$ \textbf{S} $a))$ & b \\
  \midrule
  \textsc{CoExistence}(a,b) & \textbf{F} a $\leftrightarrow$ \textbf{F} b & a,b \\
  \textsc{Succession}(a,b) & \textsc{Response}(a,b) $\land$ \newline \textsc{Precedence}(a,b) & a,b \\
    \textsc{AlternateSuccession}(a,b) & \textsc{AlternateResponse}(a,b) $\land$  \newline \textsc{AlternatePrecedence}(a,b) & a,b \\
  \midrule
  \textsc{NotCoExistence}(a,b) & (\textbf{F} a $\rightarrow \neg$\textbf{F} b) $\land$ (\textbf{F} b $\rightarrow \neg$\textbf{F} a)& a,b \\
  \bottomrule
\end{tabular}
\end{table}

In LTL, time is modeled as a linear sequence of states, or time points. At each point in time, some statements may be true. Then, LTL formulas can be used to specify the allowed behavior over the linear sequence. \textbf{F}, \textbf{X}, \textbf{G}, \textbf{U}, \textbf{S}, and \textbf{O} are LTL operators with the following meaning:  
\textbf{F}$\phi$ means that $\phi$ holds at some point in the future,
\textbf{X}$\phi_1$ means that $\phi_1$ holds in the next instant, 
\textbf{G}$\phi$ means that $\phi$ holds forever in the future,
$\phi_1$\textbf{U}$\phi_2$ means that at some point in the future $\phi_2$ will hold and until then $\phi_1$ holds;    
$\phi_1$\textbf{S}$\phi_2$ means that at some point in the past $\phi_2$ held and since then $\phi_1$ holds, and
\textbf{O}$\phi$ means that $\phi$ held at some point in the past; $\phi$, $\phi_1$, and $\phi_2$ being LTL formulas. For a formal definition of these semantics, we refer to \cite{DiCiccio2022}.

For any declarative constraint, an \emph{activation} of a constraint serves as its triggering condition. Once this condition becomes true, it expects the trace to satisfy the corresponding target condition. Conversely, if the constraint is not activated, the fulfillment of the target condition is not enforced and the constraint is (vacuously) satisfied.
For instance, $b$ is an activation for \textsc{Precedence}(a,b) and $a$ is the target because the occurrence of $b$ forces a to have occurred before. \autoref{table:declare} specifies the activation condition per template.

Given an activity sequence $\pi$ and a constraint $\phi$, assuming that both relate to the same activity set $A$, $\pi$ is said to 
satisfy $\phi$, denoted $\pi\models\phi$, if $\phi$ holds in the initial time point of the sequence. We write $\pi\not\models\phi$ if $\pi$ violates constraint $\phi$. 



\textsc{MP-Declare} extends \textsc{Declare}, among others, by introducing additional conditions on attributes of events in a trace $\sigma$, i.e., additional activation conditions.
To illustrate this concept, let us examine the constraint \textsc{Absence}(\emph{decide on application}), which is always activated at the beginning of a trace, indicating that \emph{decide on application} should never occur.
We can use an additional condition on other perspectives that must be satisfied to activate the constraint, though. For instance, we may require that when \emph{decide on application} takes place, it should be performed by a manager and otherwise should not occur. 
We can express this constraint as \textsc{Absence}(\emph{decide on application}) $\mid$ \texttt{role} $\neq$ \emph{manager}, where \texttt{role} is an event attribute and \emph{manager} its required value. Now, if (and only if) the role associated with an event is not \emph{manager}, the \textsc{Absence} constraint is activated.

\section{Framework}
\label{sec:framework}
\begin{figure}[!h]
    \centering
    \includegraphics[width=\linewidth]{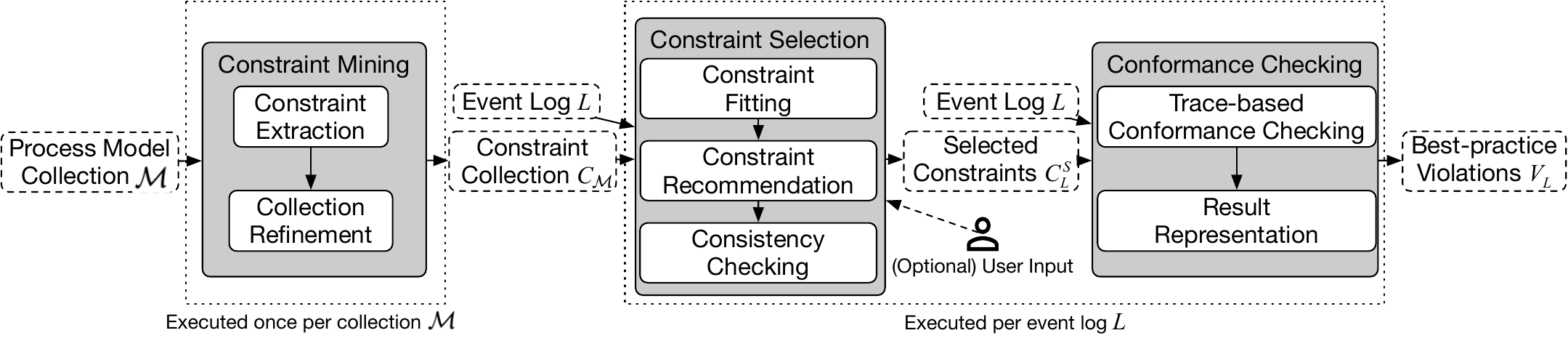}
    \caption{Framework overview.}
    \label{fig:framework}
\end{figure}

This section presents our proposed framework for detecting best-practice violations in event logs based on constraints mined from reference process models.
As illustrated in \autoref{fig:framework}, it consists of three main stages.

The input of our framework is a collection of reference process models $\mathcal{M}$. The \emph{Constraint Mining} stage extracts best practice constraints from $\mathcal{M}$, resulting in a set $\minedcoll$ containing various kinds of constraints. This stage is independent of a specific event log and, therefore, only needs to be performed once for a given collection $\mathcal{M}$.
Then, given an event log $L$, the  \emph{Constraint Selection} stage identifies which constraints from the collection $\minedcoll$ are applicable to log $L$ and fits these constraints to the log, resulting in a set of selected, fitted constraints $\instcoll^S$. This stage contains multiple sub-stages, including constraint identification and checking whether the recommended set of constraints is consistent.
Note that this stage can be executed in a fully automated manner, although users can also choose to manually refine the set $\instcoll^S$.
Finally, the \emph{Conformance Checking} stage
compares the traces in log $L$ against the constraints in $\instcoll^S$, resulting in a collection of detected best-practice violations $V_L$, presented to the user.


In the following, we continue to detail the individual steps of our framework.

\subsection{Constraint Mining}
\label{sec:framework:mining}
In the constraint mining stage, we establish a constraint collection $\minedcoll$ that represents best practices found in a reference process model collection $\mathcal{M}$. 
In the remainder of this section, we first describe the constraint types and templates that our framework covers (\autoref{sec:framework:mining:coverage}), before discussing the steps performed to mine the constraints:  \emph{Constraint Extraction} (\autoref{sec:framework:mining:extraction}) and \emph{Collection Refinement} (\autoref{sec:framework:mining:refinement}).

\subsubsection{Coverage}
\label{sec:framework:mining:coverage}

Our framework covers four types of best-practice constraints, corresponding to constraints that capture behavioral relations that should hold between activities (i.e., activity-level constraints), across business objects (i.e., inter-object constraints), within business objects (i.e., intra-object constraints), and ones that restrict the organizational roles that can perform particular process steps (i.e., role constraints). Examples of each constraint type are shown in \autoref{table:exampleconstraints}.
To represent the specific constraints, we use different sets of \textsc{Declare} templates per constraint type, as shown in \autoref{table:templateselection}.

\begin{table}[!htb]
\centering
\small
\caption{Constraint types covered by our framework with examples.}
\label{table:exampleconstraints}	
\begin{tabularx}{\linewidth}{lXX}
\toprule
\textbf{Type}& \textbf{Exemplary constraint} & \textbf{Description}\\		
\midrule
\multirow{4}{2.15cm}{Activity-level } & 
\textsc{Precedence}(\emph{Approve PO}, \emph{Process GR})& 
\emph{Approve PO} should happen before \emph{Process GR}.\\
& 
\textsc{NotCoExistence}(\emph{Reject PO}, \emph{Process GR})&
\emph{Reject PO} and \emph{Process GR} should not co-occur. \\
& 
\textsc{Succession}(\emph{Process GR}, \emph{Enter Invoice})& 
\emph{Process GR} happens if (and only if) \emph{Enter Invoice} happens later. \\
\midrule
\multirow{4}{2cm}{Inter-object} & 
\textsc{Precedence}(\emph{purchase order}, \emph{goods receipt})& 
A \emph{purchase order} should appear before a \emph{goods receipt}.\\
&\textsc{RespondedExistence}(\emph{invoice}, \emph{goods receipt}) & 
If there is an \emph{invoice} there should be a \emph{goods receipt} as well.\\
\midrule
\multirow{4}{2cm}{Intra-object} & 
\textsc{Precedence}(\emph{create}, \emph{approve}) $\mid$ \emph{purchase order}& 
For each \emph{purchase order}, \emph{create} should precede \emph{approve}.\\
& 
\textsc{AtLeastOne}(\emph{create}) $\mid$ \emph{purchase order} & 
For a \emph{purchase order}, \emph{create} should always occur.\\
\midrule
\multirow{2}{2cm}{Role } & 
\textsc{Absence}(Enter invoice) if \texttt{role} $\neq$ \emph{accounts payable}& 
\emph{Enter Invoice} should be performed by \emph{accounts payable}.\\

\bottomrule
\end{tabularx}
\end{table}

\mypar{Activity-level constraints} 
Activity-level constraints capture best practices about the behavioral relations that should hold between pairs of activities in a trace. 
In order to represent different activity inter-relations, we use \textsc{Declare} templates that capture that two activities should occur together (\textsc{RespondedExistence}, \textsc{CoExistence}) or in a certain order (\textsc{Succession}, \textsc{AlternateSuccession}),
that one activity causes (\textsc{Response}, \textsc{AlternateResponse}) or requires (\textsc{Precedence}, \textsc{AlternatePrecedence}) another activity, or that two activities should not occur together (\textsc{NotCoExistence}).

For instance, an activity-level constraint may capture that a \emph{approve purchase order} activity must precede a \emph{process goods receipt} activity (e.g., \textsc{Precedence}(\emph{approve purchase order}, \emph{process goods receipt})). 
We do not cover constraints that require activities to immediately occur after each other (e.g., \textsc{ChainResponse}) which, we assume, is too restrictive for best practices. Each such activity-level constraint is defined as follows:

\begin{definition}[Activity-level constraint]
An activity-level constraint $\minedconst \in \minedcoll$ is a tuple $\minedconst = (type, templ, $ $a_1, a_2, support)$, with  $\minedconst.type = activity$, $\minedconst.templ$ the constraint's \textsc{Declare} template, $\minedconst.a_1$ and $\minedconst.a_2$ the activities, and $\minedconst.support$ the constraint's support.
\end{definition}

Note that the constraints of all types we cover have a $support$ that indicates how often a constraint was extracted from the model collection $\mathcal{M}$ and is set when refining the collection of mined constraints $\minedcoll$.

\begin{table}[!htbp]
    \centering
    \small
    \caption{Templates used per constraint type (cf. \autoref{sec:preliminaries} for their definition)}
    \label{table:templateselection}
    \begin{tabular}{ll}
        \toprule
        \textbf{Constraint types}         & \textbf{Templates} \\
        \midrule
        \multirow{3}{*}{\textbf{Activity, Inter-object}} 
          &\textsc{RespondedExistence}, \textsc{Precedence},\textsc{AlternatePrecedence}, \\
         & \textsc{Response}, \textsc{AlternateResponse}, \textsc{Succession}, \\
        &\textsc{AlternateSuccession}, \textsc{CoExistence}, \textsc{NotCoExistence}\\
        \midrule
        \multirow{3}{*}{\textbf{Intra-object}}     
        &\textsc{AtLeastOne}, \textsc{Absence}, \textsc{ExactlyOne}, 
        \textsc{RespondedExistence}, \textsc{Precedence}, \\
        &\textsc{AlternatePrecedence}, \textsc{Response}, \textsc{AlternateResponse}, \textsc{Succession}, \\
        &\textsc{AlternateSuccession}, \textsc{CoExistence}, \textsc{NotCoExistence}\\
        \midrule
        \textbf{Role}     &   \textsc{Absence} \\
        \bottomrule
    \end{tabular}
\end{table}

\mypar{Inter-object constraints} 
Inter-object constraints capture relations that should hold between pairs of business objects occurring in a process, e.g., that a trace can only contain (activities related to) an \textit{invoice} if there is also at least one activity related to a \textit{delivery}, i.e., \textsc{RespondedExistence}(\emph{invoice}, \emph{delivery}). 
As shown in \autoref{table:templateselection}, we use the same \textsc{Declare} templates to express inter-object constraints as we use for activity-level constraints, with the difference that the constraint parameters here correspond to objects, rather than activities.
Consequently, each inter-object constraint is defined as follows:

\begin{definition}[Inter-object constraint]
An inter-object constraint $\minedconst \in \minedcoll$ is a tuple $\minedconst = (type, templ, $ $o_1, o_2, support)$, with  $\minedconst.type = interobj$, $\minedconst.templ$ the constraint's \textsc{Declare} template, $\minedconst.o_1$ and $\minedconst.o_2$ its business objects, and $\minedconst.support$ its support.
\end{definition}

\mypar{Intra-object constraints} Intra-object constraints capture relations regarding the actions, i.e., state changes, that are applied to a business object, e.g., that an \emph{order} must be \emph{checked} before it is \emph{approved}, i.e.,  \textsc{Precedence}(\emph{check}, \emph{approve}) $\mid$ \emph{order}. 
We use the vertical bar ($\mid$) to denote the constraint's business object, i.e., that both actions are applied to an \emph{order}. Although it is also possible to represent such constraints at the activity level (e.g., using \textsc{Precedence}(\emph{check order}, \emph{approve order})), explicitly capturing them for a business object offers greater flexibility in generalizing across different  business objects. For instance, in this manner, constraints like \textsc{Precedence}(\emph{check}, \emph{approve}) can be more effectively extended to other contexts, such as \emph{sales orders}. This is advantageous during the fitting of constraints for a specific event log, as we show in the \emph{Constraint Selection} stage (\autoref{sec:framework:selection}).

In addition to capturing pair-wise behavioral relations between actions, we also consider situations where specific actions must or must not be performed for certain types of objects. These are captured through the unary  \textsc{AtLeastOne}, \textsc{Absence}, and \textsc{ExactlyOne} templates, which can, e.g., capture that, if a trace contains an activity related to a \textit{purchase order item}, that trace must contain an activity that creates that item, i.e., \textsc{ExactlyOne}(\emph{create}) $\mid$ \emph{purchase order item}.

Based on this, each intra-object constraint is defined as follows:

\begin{definition}[Intra-object constraint]
An intra-object constraint $\minedconst \in \minedcoll$ is a tuple $\minedconst = (type, templ,$ $ object, arity, n_1, n_2, support)$, with  $\minedconst.type = intraobj$, $\minedconst.templ$ the constraint's \textsc{Declare} template, $\minedconst.object$ its object, and $\minedconst.arity \in \{\emph{unary},\emph{binary}\}$ its arity indicating whether $\minedconst$ is a unary or binary constraint, according to $\minedconst.templ$. 
Moreover, $\minedconst.n_1$ corresponds to the constraint's first action, $\minedconst.n_2$ represents its second action (set to $\bot$ if $\minedconst.arity = \text{unary}$), and $\minedconst.support$ pertains to its support.
\end{definition}

\mypar{Role constraints} Finally, role constraints focus on the resource perspective, restricting the execution of activities to specific roles, i.e., who in an organization can perform a given process step.
These constraints can be expressed using the \textsc{Absence} template with an activation condition. 
For instance, a role constraint may capture that an \emph{approve purchase order} activity must be performed by an employee with a \emph{manager} role, written as \textsc{Absence}(\emph{approve purchase order}) $\mid$ \texttt{role} $\neq$ \emph{manager}, with \texttt{role} $\neq$ \emph{manager} as the activation condition. Each such role constraint is defined as follows:

\begin{definition}[Role constraint]
A role constraint $\minedconst \in \minedcoll$ is a tuple $\minedconst = (type, templ,$ $a, r, support)$, with  $\minedconst.type = role$, $\minedconst.templ=$\textsc{Absence}, $\minedconst.a$ the constraint's activity, $\minedconst.r$ the role in its activation condition, and $\minedconst.support$ its support.
\end{definition}

\subsubsection{Constraint Extraction}
\label{sec:framework:mining:extraction}
The goal of constraint extraction is to derive a set of constraints $C_M$ from each process model $M \in \mathcal{M}$. Using the constraint types and corresponding templates introduced in the previous section, we base the extraction of constraints on the idea of declarative constraint mining from event logs~\cite{DiCiccio2022}, where each template is instantiated with all possible parameters (or parameter combinations) before they are checked against the traces for satisfaction. 
We apply the same idea to the set of execution sequences allowed by a model, rather than to traces from an event log, with the extraction procedure depending on the constraint type:

\mypar{Activity-level constraints} 
We extract activity-level constraints from a model $M = (A, F, R, D)$ by  first instantiating potential constraints using each template from \autoref{table:templateselection} and
each pair-wise combination of activities in $A$. Then we check which of these potential constraints is satisfied by all and activated by at least one of the execution sequences in $F$.
For instance, for the \textsc{Response} template, we check for each pair ($a_1$,$a_2$) $\in A \times A$ if \textbf{G}($a_1 \rightarrow$ \textbf{F} $a_2$) (cf. \autoref{sec:preliminaries}) holds for all $\pi \in F$ and---if so---set $\minedconst.type=activity$, $\minedconst.templ=\textsc{Response}$, $\minedconst.a_1=a_1$, and $\minedconst.a_2=a_2$ and add $\minedconst$ to $C_M$.


\mypar{Inter-object constraints} 
We extract inter-object constraints by obtaining a set of allowed object sequences $F_{\emph{obj}}$ and then checking which constraints hold between pairs of objects according to $F_{\emph{obj}}$.

To obtain $F_{obj}$, we project each (activity) execution sequence $\pi \in F$ to an object sequence $\pi_{\emph{obj}}$, 
which we achieve using a semantic extraction technique~\cite{Rebmann2022} to obtain the business objects (if any) from each activity in $\pi$.  
For instance, for $\pi=\langle \emph{create order}, \emph{check order}, \emph{approve order}, \emph{ship goods} \rangle$, we get $\pi_{\emph{obj}}=\langle \emph{order}, \emph{order}, \emph{order}, \emph{goods} \rangle$. If an activity does not relate to any business object, it is not considered in the projected trace. While uncommon, in cases where an activity relates to multiple business objects, these are considered as a single business object, assuming that in a reference process model, these are then consistently referenced together. 

Afterwards, we instantiate constraint templates between possible pairs of objects $(o_1, o_2)$ occurring in $F_{\emph{obj}}$ and check them against the sequences in $F_{\emph{obj}}$, in the same manner as done for activity-level constraints.
For instance, for the binary \textsc{RespondedExistence} template, we check for each combination ($o_1$,$o_2$) with $o_1,o_2 \in O$ if \textbf{F} $o_1$ $\rightarrow$ \textbf{F} $o_2$ holds for all $\pi_{\emph{obj}} \in F_{\emph{obj}}$ and---if so---set $\minedconst.type=interobj$, $\minedconst.templ=\textsc{RespondedExistence}$, $\minedconst.o_1=o_1$, and $\minedconst.o_2=o_2$ and add $\minedconst$ to $C_M$.

\mypar{Intra-object constraints} We extract intra-object constraints for each business object that is contained in the activities in $A$, based on the actions that are applied to such an object.

To obtain the necessary information on business objects and actions, we, therefore, again use the semantic extraction technique~\cite{Rebmann2022} (as done for inter-object constraints). 
This time, we turn the output of that approach into a function \texttt{getObjectActionPairs}, which, given an activity $a$, returns a set of object-type-action pairs obtained from $a$, where each pair is given as a tuple ($o$, $n$), indicating that action $n$ is applied to object $o$.\footnote{Note that, for clarity, we here focus on situations where each activity yields a single object-action pair, although our framework can also handle activities with multiple pairs, e.g., \textit{receive and check document}.}
For instance, $\texttt{getObjectActionPairs}(\emph{approve purchase order}) = \{(\emph{purchase order}, \emph{approve})\}$.
By applying the function to each activity $a \in A$, we obtain a set $O$ of distinct business objects that appear in the activities in $A$ and a set $N_o$ of actions per business object $o \in O$, which we use to create intra-object sequences.

As shown in \autoref{algo:projectperobject}, we create a set of intra-object sequences $F_o$ for each business object $o \in O$ as follows: 
For each $\pi \in F$, we check for each activity $a_i$ in $\pi$ if $a_i$ refers to $o$ and if so only retain the action applied to $o$. 
For instance, for $\pi=\langle \emph{create order}, \emph{check order}, \emph{approve order}, \emph{ship goods} \rangle$, we get $\pi_{\emph{order}}=\langle \emph{create}, \emph{check}, \emph{approve} \rangle$. 

\begin{algorithm}
  \caption{Create intra-object sequence sets}
  \label{algo:projectperobject}
  \small 
  \hphantom{t} \textbf{Input} $F$: set of finite execution sequences; $O$: set of business objects \\
  \hphantom{t} \textbf{Output} $res$: set containing a set of action sequences per business object
  \begin{algorithmic}[1]
      \State $res \gets \{F_o  \gets \emptyset \mid o \in O\}$ \Comment{Initialize an empty set of action sequences for each business object} 
      \For{each business object $o \in O$}
       \For{each $\pi \in F$}
       \State $\pi_o \gets \langle \rangle$
        \For{each activity $a_i$ in $\pi = \langle a_1, \ldots, a_n \rangle$}
            \State $(o_{a_{i}},n_{a_{i}}) \gets \texttt{getObjectActionPairs}(a_i)$
            \If{$o = o_{a_{i}}$} 
              \State $\pi_o \gets \pi_o + \langle n_{a_{i}} \rangle$ \Comment{Retain the action applied to $o$ and append to current sequence}
            \EndIf
          \EndFor
          \State $F_o \gets F_o \cup \{\pi_o\}$
        \EndFor
      \EndFor
      \State \textbf{return} $res$ 
  \end{algorithmic}
\end{algorithm}

Based on an intra-object set of sequences $F_o$, we can then instantiate the \textsc{Declare} templates with actions in $N_o$ in the same manner as done with entire activities for activity-level constraints.
For instance, for the binary template \textsc{Precedence}, we check for each combination $(n_1,n_2) \in N_o \times N_o$ if \textbf{G}($n_2\rightarrow$ \textbf{O} $n_1$)(cf. \autoref{sec:preliminaries}) holds for all $\pi_o \in F_o$ and---if so---set $\minedconst.type=intraobj$, $\minedconst.templ=\textsc{Precedence}$, $\minedconst.arity=\text{binary}$, $\minedconst.n_1=n_1$, $\minedconst.n_2=n_2$, and $\minedconst.object=\emph{order}$ and add $\minedconst$ to $C_M$.

\mypar{Role constraints} Finally, role constraints can be extracted independently of execution sequences from the mapping $D$ of a model $M$, which maps an activity to the role that is supposed to execute it (if any). These can thus be created by instantiating the \textsc{Absence} template with a corresponding activation condition setting $\minedconst.type=role$, $\minedconst.templ=\textsc{Absence}$, $\minedconst.a=a$, and $\minedconst.r=r$ and adding $\minedconst$ to $C_M$ for each activity $a \in A$ that is mapped to a role $r \in R$ by $D$.

\medskip
\noindent
By following this extraction procedure, we obtain a set of best-practice constraints $C_M$ per model $M \in \mathcal{M}$.
Having a large collection of models and a considerable number of constraints extracted per model, we next set out to refine this collection of constraint sets.

\subsubsection{Collection Refinement}
\label{sec:framework:mining:refinement}
Having extracted independent constraint sets $\{C_M \mid M \in \mathcal{M}\}$ per reference process model, we next establish a single refined constraint collection which is meant to aggregate all individual constraint sets. 
To this end, we standardize equal (or very similar) constraints and omit redundant constraints that are subsumed by stronger ones.
 
\mypar{Standardizing constraints} First, we standardize the actions of the constraints that we extracted to obtain a single, more concise set of constraints $\minedcoll$. 
For instance, we avoid that \textsc{Precedence}(\emph{create invoice, approve invoice}) and \textsc{Precedence}(\emph{invoice created, invoice approved}), are treated as distinct constraints.
Therefore, we standardize the activities of activity-level and action(s) of intra-object constraints by turning past-tense actions, like \emph{created} and \emph{approved}, into the present-tense, i.e., \emph{create} and \emph{approve}. We also update the word ordering of activity-level constraints accordingly, e.g., turning \textit{invoice created} into \textit{create invoice}.

After this standardization step, we can then aggregate the constraints from different models setting a constraint's support to reflect from how many process models in $\mathcal{M}$ the particular constraint was extracted.

\begin{figure}[!h]
    \centering
    \includegraphics[width=0.7\linewidth]{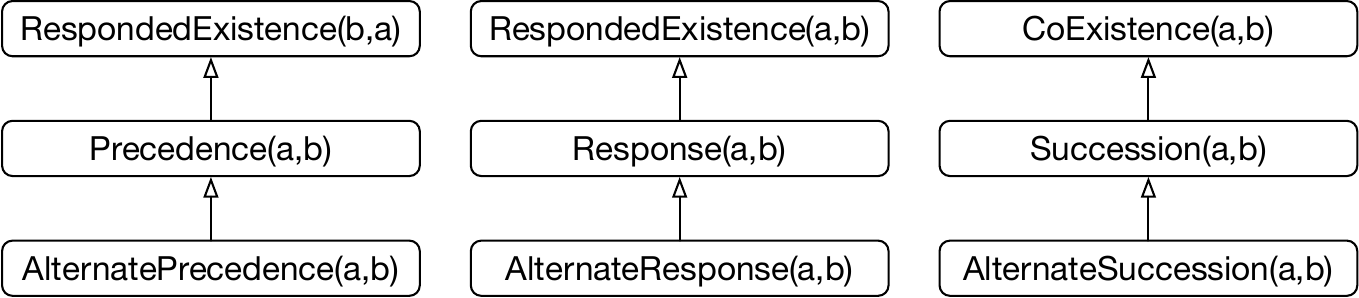}
    \caption{Subsumption relations between constraints employed by our framework (adapted from~\cite{diciccio2017}). The subsumption relation is depicted as a line starting from the subsumed template with an empty triangular arrow and ending in the subsuming one. 
    }
    \label{fig:subsumption}
\end{figure}

\mypar{Removing redundant constraints} 
Finally, we remove weaker constraints that are encompassed by stronger ones, because these weaker ones are then redundant. 
To this end, we make use of the \emph{subsumption relation} between \textsc{Declare} constraints, following the approach presented in \cite{diciccio2017}. An example would be \textsc{Response}(\emph{a, b}), which encompasses \textsc{RespondedExistence}(\emph{a, b}) ("every \textsc{Response} is also a \textsc{RespondedExistence}").
The subsumption relations between templates employed by our framework are shown in \autoref{fig:subsumption}.
For each $\minedconst^w \in \minedcoll$, we check if a stronger (subsuming) constraint $\minedconst^s$ exists that was observed the same number of times, i.e., $\minedconst^w.support = \minedconst^s.support$, and---if so---remove $\minedconst^w$ from $\minedcoll$.

\mypar{Output of the mining stage}
The output of this stage is a refined collection of mined constraints $\minedcoll$, which together with an event log $L$ serves as the input of the next stage, the \emph{Constraint Selection} stage.

\subsection{Constraint Selection}
\label{sec:framework:selection}
\begin{figure}[!h]
    \centering
    \includegraphics[width=0.7\linewidth]{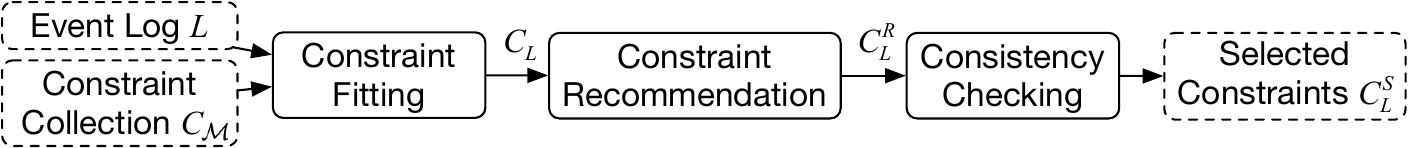}
    \caption{Overview of the constraint selection stage.}
    \label{fig:selection}
    \vspace{-1em}
\end{figure}

$\minedcoll$ is based on the reference models and therefore might contain various irrelevant constraints for a concrete event log $L$. The \textit{Constraint Selection} stage therefore aims to select those constraints from $\minedcoll$ that are most relevant for the detection of best-practice violations in a given event log $L$.
This involves three steps, as illustrated in \autoref{fig:selection}.
In the first step, we ``fit" the constraints for $L$, which yields a collection of fitted constraints $\instcoll$ (we will explain what is meant by ``fitting" below).
Out of these, we recommend the most relevant ones $\instcoll^R$, taking into account selected configuration parameters and additional user preferences.
Finally, we check the set of selected constraints $\instcoll^R$ for consistency to ensure they are without any contradictions, which yields a consistent set of selected, fitted constraints $\instcoll^S$.
In the following, we explain these steps in detail.

\subsubsection{Constraint Fitting}
\label{sec:framework:selection:instantiation}
In this step, our framework tries to fit the mined constraints in $\minedcoll$ to the contents of the given event log $L$, by relating the activities, objects, actions, and roles used in mined constraints (derived from models in  $\mathcal{M}$), to counterparts found in the event log $L$, yielding a set of fitted constraints $\instcoll$.
To do this, our framework looks for semantically similar counterparts, allowing it to, e.g., relate a 
\textsc{Precedence}(\emph{receive invoice}, \emph{pay invoice}) constraint to \textit{receive bill} and \textit{make payment} activities found in a log.
Each such fitted constraint is defined as follows:

\begin{definition}[Fitted constraint]
Given a mined constraint $\minedconst$ and an event log $L$, a fitted constraint $\instconst$ is a tuple $\instconst=(\minedconst,$ $type,$ $templ,$ $[components],$ $support,$ $sim,$ $relevance)$, with $\instconst$'s  $type$, $templ$, and $support$ being equal to those of $\minedconst$. $\instconst.[components]$ is a shorthand to refer to the list of $\instconst$'s type-specific components  (e.g., $\instconst.[components] = \instconst.a_1, \instconst.a_2$ for activity-level constraints) that are each linked to components (i.e., activities, actions, objects, or roles) found in $L$. Finally, $\instconst.sim$ is a map capturing the semantic similarity between $\minedconst$'s and $\instconst$'s components, and $\instconst.relevance$ is a score indicating the relevance of $\instconst$ for $L$.
\end{definition} 

\noindent
Creating fitted constraints for $L$ involves matching constraint components to their event log counterparts and replacing these components with their counterparts, as shown by \autoref{alg:instantiate}.
The algorithm takes as input the constraint collection $\minedcoll$ and the event log $L$, using a type-specific fitting procedure for each $\minedconst \in \minedcoll$.

\mypar{Semantic similarity computation} A key part of \autoref{alg:instantiate} is the
\texttt{match}($x_\mathcal{M}$, $x_L$) function, which is used to decide whether a constraint component $x_\mathcal{M}$ is semantically similar to an event log component $x_L$.
It is instantiated using a similarity measure \texttt{sim} and a threshold $\epsilon$, i.e., \texttt{match}($x_\mathcal{M}$, $x_L$)$:= \texttt{sim}(x_\mathcal{M},x_L) > \epsilon$. As a similarity measure, we employ the cosine similarity based on embedding vectors for $x_\mathcal{M}$ and $x_L$, which we obtain from a pre-trained \emph{sentence transformer}~\cite{reimers2019sentencebert}. These transformers are specifically designed to capture semantically meaningful representations on the level of sentences rather than individual words and can thus be applied on activities or business objects that may consist of multiple words. 
We set $\epsilon$ to 0.5 by default, ensuring that all fitted constraints are reasonably similar to the components of $L$, which avoids having to filter out a lot of constraints from $\instcoll^S$ in the next step.\footnote{Note that, in order to make \texttt{match} more or less restrictive, it is possible to use different values for $\epsilon$ depending on the type of component we aim to find matches for, i.e., setting different thresholds for activities, objects, and actions.}

\begin{algorithm}[!htb]
\caption{Constraint fitting}
\label{alg:instantiate}
\small
\hphantom{t} \textbf{Input} $\minedcoll$: set of mined constraints; $L$: event log\\
\hphantom{t} \textbf{Output} $\instcoll$: set of fitted constraints
\begin{algorithmic}[1]
    \State $\instcoll \gets \emptyset$ 
    \For{$\minedconst \in \minedcoll$}
        \If{$\minedconst.type$ is $activity$}
                        \For{$(a_{1_L}, a_{2_L}) \in A_L \times A_L$}
                            \If{$\texttt{match}(\minedconst.a_{1}, a_{1_L}) \land \texttt{match}(\minedconst.a_{2}, a_{2_L}) \land a_{1_L} \neq a_{2_L}$                             \label{line:activitymatch}}
                                \State \texttt{FitActivityConstraint}($\minedconst$, $a_{1_L}$, $a_{2_L}$)  \label{line:activityinstant}
                            \EndIf               
                        \EndFor
          \EndIf
        \If{$\minedconst.{type}$ is $interobj$}
			  \For{$(o_{1_L}, o_{2_L}) \in O_L \times O_L$}
			\If{$\texttt{match}(\minedconst.o_{1}, o_{1_L}) \land \texttt{match}(\minedconst.o_{2}, o_{2_L}) \land o_{1_L} \neq o_{2_L}$                             \label{line:objectmatch}}
			\State \texttt{FitInterObjectConstraint}($\minedconst$, $o_{1_L}$, $o_{2_L}$)  \label{line:objectinstant}
			\EndIf               
			\EndFor
        \EndIf
        
              \If{$\minedconst.type$ is $intraobj$}
			            \For{$ o_L \in  O_L$}
			\If{$\texttt{match}(\minedconst.object, o_L)$} \label{line:intramatchobj}
              \If{$\minedconst.arity=$ unary}
                	\For{$n_{L} \in N_{o_L}$}
                		\If{$\texttt{syn}(\minedconst.n_1, n_{L})$}
                			\State \texttt{FitIntraObjectConstraint}($\minedconst$, $n_{L}$, $\bot$)
                		\EndIf
                	\EndFor
                \EndIf

                \If{$\minedconst.arity=$ binary}
                 \For{$(n_{1_L}, n_{2_L}) \in N_{o_L} \times N_{o_L}$}
			 \If{$\texttt{syn}(\minedconst.n_{1}, n_{1_L}) \land \texttt{syn}(\minedconst.n_{2}, n_{2_L}) \land n_{1_L} \neq n_{2_L}$}
              \State \texttt{FitIntraObjectConstraint}($\minedconst$, $n_{1_L}$, $n_{2_L}$)
              \EndIf
              \EndFor
              \EndIf  
              \EndIf
              \EndFor
        \EndIf
        
        \If{$\minedconst.type$ is $role$}
            \For{$a_L \in A_L$}
                \If{$\texttt{match}(\minedconst.a, a_L)$} \label{line:role:actmatch}
                    \For{$r_L \in R_L$}
                        \If{$\texttt{match}(\minedconst.r, r_L)$} 
                            \State \texttt{FitRoleConstraint}($\minedconst$, $a_L$, $r_L$) \label{line:role:roleinstant}
                        \EndIf
                    \EndFor
                \EndIf
            \EndFor
        \EndIf
    \EndFor
    \State \Return $\instcoll$
\end{algorithmic}
\end{algorithm}

\mypar{Activity-level constraints} 
For an activity-level constraint $\minedconst$, we need to find activities in $L$ that correspond to the constraint's activity components, $\minedconst.a_{1}$ and $\minedconst.a_{2}$.
Specifically, we create a fitted version of $\minedconst$ for any pair of distinct log activities $a_{1_L}$ and $a_{2_L}$ for which both \texttt{match}($\minedconst.a_{1}, a_{1_L}$) and \texttt{match}($\minedconst.a_{2}, a_{2_L}$) hold (lines~\ref{line:activitymatch}--\ref{line:activityinstant}).

It is important to note that this can result in multiple fitted constraints being added to $\instcoll$ that stem from a single mined constraint $\minedconst$. 
To illustrate this, consider a mined constraint that specifies that requests must first be examined before a decision can be made on them, i.e., \textsc{Precedence}(\emph{examine request}, \emph{decide on request}). It is well-imaginable that a process like this actually uses different kinds of examinations for different types of cases or situations, e.g., having \textit{perform basic examination} and \textit{perform thorough examination}. In such a scenario, the mined constraint needs to be fitted twice, so that both types of examinations are covered, yielding constraints capturing \textsc{Precedence}(\emph{perform simple examination}, \emph{decide on request}) and \textsc{Precedence}(\emph{perform thorough examination}, \emph{decide on request}).

\mypar{Inter-object constraints}
To be able to fit an inter-object constraint $\minedconst$, we need to find two business objects in  $L$ that match the business objects of $\minedconst$.
To do this, we use $O_{L}$ to denote the set of business objects in $L$, which is obtained using \texttt{getObjectActionPairs} (see \autoref{sec:framework:mining:extraction}), i.e., $O_{L}=\{o \mid (o,n) \in \{\texttt{getObjectActionPairs}(a) \mid a \in A_L\} \}$. 
Then, the fitting procedure for inter-object constraints is identical to the procedure for activity-level constraints, with the exception that we match the constraint's objects to objects in $O_L$ (lines~\ref{line:objectmatch}--\ref{line:objectinstant}).

\mypar{Intra-object constraints}
For an intra-object constraint $\minedconst$, we need to find both a matching business object and one or two corresponding actions (depending on if $\minedconst$ is a binary or unary constraint) in $L$.
To this end, we first check for a matching business object $o_L$, in the same manner as done for inter-object constraints (line~\ref{line:intramatchobj}). If found, we next check for matches between $\minedconst$'s action(s) and those in the log that are applied to $o_L$, i.e., in the set $N_{o_L}$ (see \autoref{sec:framework:mining:extraction}).

A distinction we make here is that we do not look for corresponding actions based on semantic similarity (i.e., using \texttt{match}), but rather look for synonyms of actions. 
We make this distinction because actions generally correspond to verbs, which is a specific class of words that is covered well by lexical resources, such as \emph{WordNet}~\cite{wordnet}, which provide curated lists of similar terms (e.g., synonyms). Therefore, by checking for synonymous actions, we can use a restrictive, but highly precise matching strategy for actions.\footnote{Note that this synonym-based matching strategy using \texttt{syn} can be straightforwardly replaced with
	\texttt{match} if desired.}

For instance,  consider a constraint $\minedconst$ corresponding to \textsc{AtLeastOne}(\emph{check}) $\mid$ \emph{order}  and an event log that relates to \textit{purchase order} objects, including an \textit{examine purchase order} activity.
Then, our framework would recognize that \textit{order} and \textit{purchase order} are semantically similar, i.e., \texttt{match}(\textit{order}, \textit{purchase order})$=true$, and that \textit{examine} is a synonym of \textit{check}, i.e.,  \texttt{syn}(\textit{check}, \textit{examine})$=true$, yielding a fitted constraint $\instconst$ corresponding to \textsc{AtLeastOne}(\emph{examine}) $\mid$ \emph{purchase order}.

\mypar{Role constraints}
Finally, to fit a role constraint $\minedconst$, we need to identify both a matching activity and a matching role in $L$.
To this end, we first check for a matching activity $a_L \in A_L$ in the same manner as for activity-level constraints (line \ref{line:role:actmatch}).
If found, we next check for matches between $\minedconst$'s role $\minedconst.r$ (captured in its activation condition) and the roles in the log, i.e., in the set $R_L$. If for $r_L \in R_L$, \texttt{match}($\minedconst.r$, $r_{L}=true$) holds, we fit $\minedconst$ with $a_L$ and $r_L$ (line \ref{line:role:roleinstant}).

\medskip
\noindent
When fitting a constraint, we populate $\instconst$'s similarity map $\instconst.sim$
such that it maps the matched components from $\minedconst$ and their counterparts in $\instconst$ to their respective similarity scores, which we will use for recommendation. Having iterated through all constraints $\minedconst \in \minedcoll$, the algorithm outputs the set of fitted constraints $\instcoll$.

\subsubsection{Constraint Recommendation} 
While from the previous step we have obtained a set of fitted constraints, this set might still be too large, or contain constraints that are uninteresting. In the recommendation step, we therefore aim to select a subset of constraints from $\instcoll$ that are most relevant to the event log $L$.
To this end, we first compute a relevance score for each constraint, which we then use to select a set $\instcoll^R$ of recommended constraints, according to user-defined parameters regarding size and relevance.

\mypar{Relevance computation}
\label{sec:framework:selection:relevance}
To be able to recommend appropriate constraints, we need an indicator of how relevant a constraint $\instconst \in \instcoll$ is to an event log $L$.
To achieve this, we compute a relevance score based on $\instconst$'s semantic similarity scores and support. We consider these two aspects, since a  higher semantic similarity indicates that a constraint is more applicable to the specific context of the given event log $L$ (e.g., because the activities from the original, mined constraint are highly similar to those found in $L$), whereas higher support indicates that a constraint is more generally applicable 
to processes (because it was found in a larger number of models in $\mathcal{M}$). We capture these aspects as follows:

\medskip
\noindent
$\instconst.relevance = \omega * \texttt{avg}(\{s \mid ((x,y),s) \in \instconst.sim\}) + {(1-\omega)} * (\frac{\instconst.support}{\texttt{max}(\{\instconst'.support \mid \instconst' \in \instcoll, \instconst'.type = \instconst.type\} })
)$
\medskip

As shown, we combine the average similarity score $\texttt{avg}(\{s \mid ((x,y),s) \in \instconst.sim\})$ observed for the constraint with the constraint's support $\instconst.support$, normalized by dividing it by the maximum support of a constraint of the same type in $\instcoll$, such that both, similarity and support, have a comparable range.
The two aspects can be weighed differently by adapting $\omega \in [0, 1]$, e.g., to give a higher weight to support.

\mypar{Constraint selection}
The computed relevance scores can be employed to obtain a selection of recommended constraints $\instcoll^R$. Our framework allows users to select all constraints from $\instcoll$ that have a relevance higher than a certain threshold $\tau$, or to select the top-$k$ constraints from $\instcoll$ that have the highest relevance scores, thus providing users control over either the minimum desired relevance score $\tau$ or the maximal number of constraints $k$.
Users can also apply the selection strategies in a type-specific manner, i.e., by specifying different relevance thresholds $\tau_{type}$ for each constraint type, or by selecting the top-$k$ constraints per type.

\mypar{User input} 
After providing a set of recommended constraints $\instcoll$, the user can decide if they continue without intervention or if they want to inspect the recommended constraints and optionally filter the collection manually. 
For instance, if they find that some recommended constraints involve an object that is not interesting from a compliance perspective, they can remove all constraints that involve that object.

\subsubsection{Consistency Checking} 
The final step of the \emph{Constraint Selection} stage ensures that $\instcoll^R$ contains no inconsistencies. 
For instance, we want to avoid that the set contains both \textsc{CoExistence}(a,b) and \textsc{NotCoExistence}(a,b) constraints, since these directly contradict each other. Note that inconsistencies can also be more complex due to transitivities, e.g., a set of constraints \textsc{Response}(a,b), \textsc{Response}(b,c) and \textsc{NotCoExistence}(a,c) is also inconsistent, since the first two constraints transitively state that $a$ should eventually be followed by $c$, whereas the final constraint explicitly forbids this.
Since such combinations are hard, if not impossible, to spot for humans \cite{DBLP:conf/ecis/NagelD22}, automated consistency checking and resolution is required.

To operationalize this, we use an existing approach for inconsistency resolution in declarative process specifications~\cite{corea2021interactive}. Given a set of constraints $C$, this approach identifies a set of minimal correction sets \textsf{MCS}($C$), which are sets of constraints that, if removed from 
$C$, make $C$ consistent:

\begin{definition}[Minimal correction sets]
 Given a constraint set $C$, a set $\Phi\subseteq C$
is a minimal correction set of $C$, if $C\setminus\Phi$ is consistent, and $\forall \Phi'\subset\Phi: C\setminus\Phi'$ is inconsistent. We denote \textsf{MCS}(C) 
as the set of minimal correction sets for the constraints in C.
\end{definition}

If the inconsistency resolution approach detects inconsistencies in $\instcoll^R$, 
our framework, by default, deletes the constraints from the correction set in \textsf{MCS}($\instcoll^R$) with the lowest cumulative relevance score.
This ensures that we obtain a consistent set $\instcoll^S$ of constraints, while maximizing the overall relevance of the retained constraints.
If there are no inconsistencies, $\instcoll^R$ simply becomes $\instcoll^S$.

\mypar{User input}
The employed inconsistency resolution approach~\cite{corea2021interactive} also allows a user to select specific constraints that should not be removed from $\instcoll^R$, giving them control over the recommended correction sets. Furthermore, the user has the option to inspect the minimal correction sets and choose to remove one of these based on their preferences instead of automatically having the one removed that maximizes relevance.

\mypar{Output of the selection stage}
The output of this stage is a collection of fitted constraints $\instcoll^S$ that is free of inconsistencies. $\instcoll^S$ is used in the subsequent stage to check for best-practice violations.

\subsection{Conformance Checking}
Our framework's final stage uses the set of selected constraints $\instcoll^S$ to check for best-practice violations in the event log $L$ and, if found, presents these to the user.
To this end, the framework first identifies such violations for each trace, yielding a set of violations $V_L$.  Afterwards, it aggregates the violations in $V_L$ to the log-level and adds an explanation of each type of violation.

\subsubsection{Trace-based Conformance Checking}
The goal of this step is to check which constraints are violated per trace $\sigma \in L$. 
To this end, our framework creates a collection $V_L$  of trace-constraint pairs ($\sigma$,$\instconst$), where each pair represents the violation of $\instconst$ in $\sigma$. 
To establish these pairs, our framework checks per trace $\sigma \in L$ and each constraint $\instconst \in \instcoll^S$ if that constraint is violated, i.e., if $\sigma \not\models \instconst$. 
The checking process is analogous to the one used in constraint mining (cf. \autoref{sec:framework:mining}). However, while in constraint mining we assess whether a constraint is fulfilled across all sequences allowed by a model, here we focus on identifying specific traces where a constraint is not met. 
Given a trace $\sigma$, the conformance checks per constraint type are performed as follows:


\mypar{Activity-level and role  constraints}
For an activity-level or a role constraint $\instconst$, our framework can directly check if $\sigma \not\models \instconst$ holds and, if so, add ($\sigma$, $\instconst$) to $V_L$.

\mypar{Inter-object constraints}
For an inter-object constraint $\instconst$, our framework checks against the object sequence $\sigma^a_{obj}$ that it obtains by projecting the activity sequence $\sigma^a$ to its objects, as also done when mining constraints of this type.
Then, it checks if $\sigma^a_{obj} \not\models \instconst$ holds and, if so, adds ($\sigma$, $\instconst$) to $V_L$.

\mypar{Intra-object constraints} 
For an intra-object constraint $\instconst$, the framework checks against the action sequence $\sigma_o$ for its business object $o=\instconst.object$, which it obtains in the same manner as done in \autoref{algo:projectperobject}. 
Then, it checks if $\sigma_o \not\models \instconst$ holds and, if so, adds ($\sigma$, $\instconst$) to $V_L$.

\subsubsection{Result Representation}
In this final step, our framework creates a more concise and informative summary of the violations for the user by aggregating them across traces.
Specifically, it groups violations per constraint type and template and adds a textual explanation to each such violation. An example of this is shown in \autoref{table:exampleaggoutobj}. To generate the explanations for the violations, we use a standard sentence template for each combination of constraint type and \textsc{Declare} template.

\begin{table}[!htbp]
    \centering
    \small
    \begin{tabularx}{\linewidth}{lXr}
        \toprule
        \textbf{Constraint} & \textbf{Constraint explanation} & \textbf{Violating traces} \\
        \midrule
         \textsc{AtLeastOne}(\emph{create})$|$\emph{order} &  Each \textit{order} must be \textit{created} & $\{\sigma_6,\dots,\sigma_{76}\}$ \\
        \textsc{Precedence}(\emph{check}, \emph{approve})$|$\emph{invoice} &  An \textit{invoice} must be \textit{checked} before it is \textit{approved} & $\{\sigma_{23}\}$ \\
         \bottomrule
    \end{tabularx}
    \caption{Exemplary result representation of intra-object constraint violations.}
    \label{table:exampleaggoutobj}
\end{table}

\section{Evaluation}
\label{sec:evaluation}
In this section, we aim to demonstrate the effectiveness of our framework to select relevant constraints for detecting best-practice violations. 
To do so, we test if the constraints that our framework selects for a given log can be used to find known violations. 
We describe the data collection in \autoref{sec:data} and the experimental setup in \autoref{sec:setup}. 
In \autoref{sec:results}, we present our evaluation results demonstrating our approach's efficacy in identifying best-practice violations and that it outperforms a baseline approach in both scope and accuracy. 
The implementation, data collection, evaluation pipeline, and raw results are all
available in our repository\footnote{\url{https://github.com/a-rebmann/semantic-constraint-miner/}}.

\subsection{Data Collection}
\label{sec:data}
Our goal is to show that our framework is effective in selecting constraints that are relevant for a given event log and that these constraints can be used to identify best-practice violations. 
Therefore, we need a process model collection to extract constraints from and event logs with known best-practice violations, i.e., with behavior that deviates from a process model that represents desired behavior.

\mypar{Process model collection}
Since there is no public reference model collection available, we use a collection of real-life process models. We use this collection for both extracting constraints and generating event logs with violations. For the experiments, we apply a cross-validation procedure as explained later.
Specifically, we use the public \emph{SAP-SAM} data set~\cite{sola2023sap}, a large process model collection created by academic users of a commercial process modeling tool. 

Given that there is no quality assurance for the entire model collection, we select only English models that fulfill a set of requirements to reduce data quality issues. In particular, each model needs to have between 5 and 50 elements, pass the BPMN syntax check (using the functionality of a commercial process modeling tool), and must be transformable into a sound workflow net. The former two requirements reduce the probability that models with barely any (too few elements) or pointless (too many elements and syntax errors) behavior are included, whereas the latter ensures that we can generate proper event logs from the models.
From those we randomly select 1,500 models, primarily in order to keep run times manageable.
Note that because this is not a real \emph{reference} process model collection, we assume that the behavior of the contained models captures best practices.

\mypar{Generating event logs with violations}
To obtain logs with known violations from this process model collection, we play out the models and insert noise:
For each model, we use its workflow net to generate an event log that contains a single trace of each variant that is possible in the net (in case of loops, each loop is executed at most once). For nets with fewer 
than 100 variants, we keep playing out traces until the log has 100 traces, so that we have a minimum amount of traces available per log for noise insertion.
Then, to add violations to the logs, we select traces to introduce noise into, with a probability of 50\% per trace.
If a trace is chosen for noise insertion, we either randomly add, remove, or swap events or assign a different role to an event than the original one. After performing a noise-insertion action for a selected trace, there is a 50\% probability to insert noise again (repeated until false).
The characteristics of the log collection (before and after noise insertion) obtained in this manner are shown in \autoref{table:data}. As depicted there, the complexity of the logs varies considerably. For instance, the logs have 5 activities on average, whereas the maximum number observed is 18 and the number of object types is 3.9 on average, whereas the maximum is 17. 
The differences between the original and noisy logs become clear when looking at the variants. While the original logs contain 3.5 different variants on average, this increases to 30.9 for the noisy logs.

\begin{table}[!htb]
    \centering
    \caption{Characteristics of the original and noisy logs.}
    \label{table:data}
    \small
    \begin{tabular}{llcccccS[table-format=6.0]cc}
        \toprule
        \textbf{Collection} & \textbf{Count} & \multicolumn{2}{c}{\textbf{Activities}} &
        \multicolumn{2}{c}{\textbf{Object types}} & 
         \multicolumn{2}{c}{\textbf{Variants}} & \multicolumn{2}{c}{\textbf{Variant length}}   \\
        && \textbf{avg.} & \textbf{max.} 	& \textbf{avg.} & \textbf{max.}		& \textbf{avg.} & \textbf{max.}  & \textbf{avg.} & \textbf{max.} \\
        \midrule
        
        Original logs & 1,500 & 4.9 & 18 &  3.9 & 17 & 3.5 & 720 & 3.7 & 10\\
        Noisy logs & 1,500  & 4.8 & 18 &  3.9 & 17 &  30.9 & 703 & 3.9 & 14 \\
        \bottomrule
    \end{tabular}
\end{table}

The inserted noise then leads to violations of activity, inter-object, intra-object, and role constraints.
Because we assume that a process model established by users contains best-practice behavior, any behavior not allowed by the model can be considered a best-practice violation. Hence, we compute the known violations for an event log $L$ by verifying for each trace $\sigma \in L$ if $\sigma$ violates the behavior of the model $M$ that was used to generate $L$ and recording violations of this behavior in a collection of true violations $L^M_V$. 
Statistics about the violations contained in our log collection are depicted in \autoref{table:trueviolations}. As shown there, on average 40\% of traces are affected by activity-level violations, 28\% by inter-object violations, 38\% by intra-object violations, and 9\% by role violations. Across constraint types, the average number of violations per log is considerably higher than the number of affected traces, indicating that frequently there are multiple violations per trace.

\begin{table}[!htb]
    \centering
    \caption{Violations per noisy log.}
    \label{table:trueviolations}
    \small
    \begin{tabular}{lrrrr}
    \toprule
     \textbf{Const. Type} & \multicolumn{2}{c}{\textbf{Affected traces}} & 
    \multicolumn{2}{c}{\textbf{Violations}} \\
    &\textbf{avg.} & \textbf{max.} & \textbf{avg.} & \textbf{max.} \\
\midrule
\textbf{Activity} & 40.77 & 312 & 169.00 & 1979 \\
\textbf{Inter-object} & 28.34 & 276 & 95.35 & 1453 \\
\textbf{Intra-Object} & 38.31 & 199 & 91.22 & 905 \\
\textbf{Role} & 9.14 & 150 & 11.47 & 172 \\
\bottomrule
\end{tabular}
\end{table}

\subsection{Setup}
\label{sec:setup}
\mypar{Implementation}
We implemented our framework and evaluation pipeline in Python. 
For handling the import and generation of event logs, we used PM4Py~\cite{berti2019process} and as a basis for constraint checking, we used Declare4Py~\cite{donadello2022declare4py}. 
To assess semantic similarity between activities and business objects, we generated sentence transformer embeddings~\cite{reimers2019sentencebert} using the \emph{all-MiniLM-L6-v2} pretrained language model\footnote{Available here: \url{https://huggingface.co/sentence-transformers/all-MiniLM-L6-v2}}. 
Finally, as described in \autoref{sec:framework}, we use existing approaches~\cite{Rebmann2022,corea2021interactive}
to extract objects and actions from activities and to check for inconsistencies in constraint sets.

\mypar{Configurations}
We test various settings for the relevance threshold $\tau$, the weight of the semantic similarity $\omega$ when computing the relevance of a constraint, and $k$ that determines how many of the most relevant constraints per type are selected. In particular, we run experiments for $\tau \in \{0.5, 0.8\}$ to investigate the effect of a low vs. a high relevance score, $\omega \in \{0.5, 0.9\}$ to investigate the importance of support, and $k\in \{10, 100, 250\}$ to investigate how the number of selected constraints impacts performance.

\mypar{Baseline}
We compare our framework to the semantic-anomaly-detection approach proposed by Van der Aa et al.~\cite{VanDerAa2021}. 
Its goal is to detect whether behavior recorded in an event log does not make sense from a semantic point of view. 
Therefore, it can also be used to detect violations of general behavioral relations captured in models of well-engineered processes, i.e., reference process models, which makes it a suitable baseline for our framework.
Their approach populates a knowledge base with semantic relations that reflect appropriate process executions. The knowledge records for this population procedure are extracted from (1) a general-purpose linguistic resource that captures relations between verbs and (2) a process model collection (i.e., the same process model collection we use to mine constraints from). 
For a given trace, the approach checks whether pairs of actions that are applied to the same object violate records in the knowledge base, indicating semantic anomalies. 
We compare our work against the configuration with the best reported results in the original paper, referred to as \textit{SEM4} in their experiments. Importantly, this configuration leverages semantic similarity to enhance the generalizability of the rules stored in the knowledge base.
It is worth stressing that, because of its focus on actions applied to the same object, the baseline can solely detect intra-object constraint violations.

\mypar{Cross validation}
Because we use the same model collection $\mathcal{M}$ for constraint mining and detection, we conduct a 5-fold cross validation. To this end, we randomly split the model collection into 5 sets and use models of 4 sets for constraint mining and the noisy logs generated from the 5th set for testing the violation detection. We run the experiments 5 times, so that each set acts as the test set once.

\mypar{Measures} 
We check for violations in the noisy log $L^M$, given the original model $M$, which yields a collection of detected violations $L^M_D$. 
For each detected violation $v \in L^M_D$, we check if $v$ is contained in the set of true violation $L^M_V$ (computed as explained above) resulting in a true positive (TP) or false positive assessment (FP). Any violation that is present in $L^M_V$ but has not been detected by our framework, i.e., is not included in $L^M_D$, is deemed a false negative (FN). Using this approach, we quantify the precision, given by TP/(TP+FP) and the recall, given by TP/(TP+FN).
Beyond these, we report on run time for the different stages.

\subsection{Results}
\label{sec:results}
We first report on the constraints mined from the model collection, before describing the overall results of the violation detection. 
Then, we report on the results per constraint type and inspect the detailed results of selected logs to provide insights into when our framework performs well and when it fails. 
Finally, we compare the performance of our framework against the baseline and reflect on its run time.

\mypar{Extracted constraints}
Our framework extracted more than 250,000 constraints from the 1,500 process models. As shown in \autoref{table:evalconsts}, this includes more than 150,000 activity-level, almost 60,000 inter-object, more than 42,000 intra-object constraints, and more than 3,000 role constraints. Per process model, it extracted at most 662 activity-level constraints (104.0 on average), 416 inter-object constraints (54.1 on average), 366 intra-object constraints (33.3 on average), and 17 role constraints (6,3 on average).

\begin{table}[!h]
    \centering
    \caption{Characteristics of the constraints extracted from the model collection before collection refinement}
    \label{table:evalconsts}
    \small
    \begin{tabular}{lrcc}
        \toprule
        \textbf{Const. type}         & \textbf{Count} & \textbf{Avg. per model} & \textbf{Min./Max. per model} \\
        \midrule
        \textbf{Activity}     &     150,082  &        104.0        &           2   /        662      \\
        \textbf{Inter-object}&     59,452  &          54.1      &            2    /         416     \\
        \textbf{Intra-object} &     42,377  &          33.3      &            3    /          366     \\
        \textbf{Role}     &       3,349&           \hphantom{5}6.3     & 0  / 17 \\
        \bottomrule
    \end{tabular}
\end{table}

After refinement, i.e., after equal or highly similar constraints were standardized and weaker, redundant ones were filtered out, 55,913 activity-level, 20,920 inter-object, 8,223 intra-object constraints, and 2,387 role constraints remain, as depicted in \autoref{table:evalconstsagg}. 
Many constraints were extracted from multiple models, as shown by the number of constraints with a support of $>1$. Notably, more than half of the intra-object constraints (4,484) were extracted from multiple models, which suggests that these are more generally applicable than the other types of constraints, for which this ratio is lower.

\begin{table}[!h]
    \centering
    \caption{Characteristics of the constraint collection after aggregation and refinement. 
    }
    \label{table:evalconstsagg}
    \small
    \begin{tabular}{lrrr}
        \toprule
        \textbf{Const. type}         & \textbf{Count} &  \textbf{Support $>1$} \\
        \midrule
        \textbf{Activity}     &  55,931    &    5,277    \\
        \textbf{Inter-object} &  20,920    &    2,965    \\
        \textbf{Intra-object} &  8,223     &    4,484   \\
        \textbf{Role}         &  2,387     &      266   \\
        \bottomrule
    \end{tabular}
\end{table}

\mypar{Overall detection results}
\autoref{table:overallresults} depicts the evaluation results per configuration for detecting best-practice violations from the noisy logs, averaged across logs. Our experiments showed that the performance trends per configuration are always the same across folds. Moreover, we found that weighing the semantic relevance $\omega$ with 0.9 for all constraint types consistently yields slightly better results than weighing the support equally. For the other parameters the variety of the performance achieved is more substantial, though. Therefore, we report on results with $\omega=0.9$ here for brevity.\footnote{Detailed results including all configurations can be found in our repository.} 

\begin{table}[!t]
    \centering
        \small
    \caption{Results of detecting best-practice violations per constraint type for various configurations (averaged across logs). We generally favor high recall over high precision and the best trade-off between recall and precision per constraint type is printed in bold. $k$ determines the number of most relevant constraints selected per type. $\tau$ is a threshold for the minimum relevance score that a constraint must have to be selected. The weight factor of the relevance score, $\omega$, is set to 0.9 in all configurations.}
    \label{table:overallresults}
    \begin{tabular}{lrrrrrccc}
        \toprule
        \textbf{Type} & \multicolumn{2}{c}{\textbf{Config.}} & \textbf{TP} &  \textbf{FP} & \textbf{FN} &  \textbf{Precision} & \textbf{Recall}  \\
        & $k$ & $\tau$ & & & & &\\
        \midrule
        \multirow{6}{*}{\textbf{Activity}}
        & 10 & 0.5 & 77.10 & 62.04 & 91.90 & 0.70 & 0.51 \\
        

        & 100 & 0.5 & 134.22 & 230.83 & 34.78 & \textbf{0.43} & \textbf{0.82} \\

        & 250 & 0.5 & 139.80 & 286.17 & 29.20 & 0.39 & 0.84 \\

        \cline{2-8}
        \rule{0pt}{10pt}
        & 10 & 0.8 & 54.54 & 24.82 & 114.46 & 0.86 & 0.35 \\
        
        
         & 100 & 0.8 & 80.08 & 76.61 & 88.92 & 0.75 & 0.49 \\
         
         & 250 & 0.8 & 80.27 & 79.64 & 88.73 & 0.74 & 0.49 \\

         
        \midrule
        \multirow{6}{*}{\textbf{Inter-object}}
        & 10 & 0.5 & 50.56 & 77.46 & 44.79 & 0.59 & 0.58 \\
        

        & 100 & 0.5 & 77.57 & 231.39 & 17.77 & \textbf{0.34} & \textbf{0.82} \\

        & 250 & 0.5 & 79.47 & 277.79 & 15.88 & 0.32 & 0.84 \\

        \cline{2-8}
        \rule{0pt}{10pt}
        & 10 & 0.8 & 42.07 & 52.11 & 53.28 & 0.70 & 0.48 \\
        

        & 100 & 0.8 & 53.87 & 98.07 & 41.48 & 0.59 & 0.59 \\

        & 250 & 0.8 & 53.87 & 98.33 & 41.48 & 0.59 & 0.59 \\

        \midrule
        \multirow{6}{*}{\textbf{Intra-object}}
        & 10 & 0.5 & 51.69 & 26.46 & 39.53 & 0.84 & 0.70 \\
        
        
        & 100 & 0.5 & 68.64 & 131.71 & 22.58 & \textbf{0.72} & \textbf{0.79} \\

        & 250 & 0.5 & 69.55 & 172.97 & 21.67 & 0.70 & 0.79 \\

        \cline{2-8}
        \rule{0pt}{10pt}
        & 10 & 0.8 & 46.74 & 21.26 & 44.48 & 0.86 & 0.59 \\
        

        & 100 & 0.8 & 58.47 & 94.93 & 32.75 & 0.78 & 0.65 \\

        & 250 & 0.8 & 58.80 & 104.46 & 32.42 & 0.77 & 0.65 \\

        \midrule
        \multirow{6}{*}{\textbf{Role}}
        & 10 & 0.5 & 6.51 & 16.73 & 4.96 & 0.60 & 0.57 \\
        

        & 100 & 0.5 & 9.01 & 37.39 & 2.46 & \textbf{0.49} & \textbf{0.80} \\

        & 250 & 0.5 & 9.08 & 38.31 & 2.39 & 0.49 & 0.80 \\

        \cline{2-8}
        \rule{0pt}{10pt}
        & 10 & 0.8 & 4.39 & 8.47 & 7.08 & 0.78 & 0.38 \\
        

        & 100 & 0.8 & 4.77 & 10.78 & 6.69 & 0.76 & 0.41 \\

        & 250 & 0.8 & 4.77 & 10.78 & 6.69 & 0.76 & 0.41 \\

        
        \midrule
        \multirow{1}{*}{\textbf{Overall}}
        & 100  & 0.5 & 72,36 & 157,82&  19.40& 0.50 & 0.81 \\
        \bottomrule
    \end{tabular}
\end{table} 

Overall, we find that both precision and recall vary considerably across configurations (0.37--0.86 resp. 0.35--0.85). Because we generally favor high recall over high precision when detecting best-practice violations, the best out of the tested configurations, which achieves good recall while maximizing precision is $k=100$ combined with a low relevance threshold ($\tau=0.5$) for all constraint types.
By picking this configuration, we achieve an average precision of 0.50 and a recall of 0.81 across constraint types.\footnote{Note that, while this configuration achieves the best results across constraint types, we are not bound to using the same configuration per type.}
Selecting a high relevance threshold ($\tau=0.8$) yields considerably lower recall scores (0.35--0.65) compared to when using a lower threshold ($\tau=0.5$; 0.57--0.85). This indicates that applying such a high relevance threshold is too restrictive, causing a considerable amount of violations to remain undetected.

\begin{figure}[!t]
  \centering
    \centering
    \includegraphics[width=0.6\textwidth]{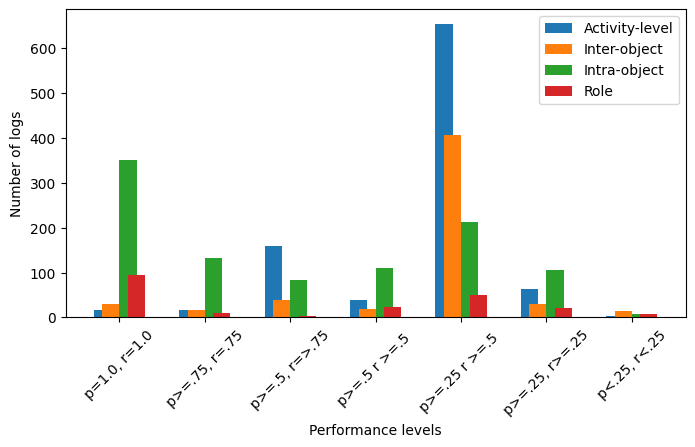}
  \caption{Number of logs for different performance levels per constraint type (considering logs for which there were violations to be detected for the given type of constraint); p=precision, r=recall, lower performance levels exclude logs that are already included in higher ones.}
  \label{fig:performance}
\end{figure}

We also looked into per-log performance to get insights into how often our framework achieves excellent, good, and poor performance.
\autoref{fig:performance} shows the number of logs for which various performance levels are reached, using the best configuration per constraint type ($k=100$, $\tau=0.5$). We find that our framework achieved  perfect precision and recall for a considerable number of logs. Especially for intra-object constraints this was the case for more than 350 logs. For the other types the framework does not perform as well but still for the vast majority, i.e., more than 90\% of the logs, a recall (which is the most important score) of more than 0.5 was achieved. Only for very few logs both precision and recall are poor, i.e., both below 0.25. 

\mypar{Results per constraint type}
To get a better understanding of when the framework excels and when it fails, we next discuss the results per constraint type by analyzing the impact of the parameter settings on performance and inspecting the results achieved for individual event logs.

\mypartwo{Activity-level constraints}
For activity-level constraints, the best configuration achieves a moderate precision of 0.43 and a good recall of 0.82.
We find that we primarily obtain perfect scores for rather standard types of processes, where organization-specific deviations are expected to occur rather rarely. For instance, for an order-handling event log, all 246 violations were correctly identified without any false positives, recognizing, e.g., that \emph{confirm order} must precede \emph{emit invoice}, that \emph{archive order} should only happen after \emph{receive payment}, and that \emph{ship product} should not occur before \emph{get shipment address} if the latter occurs. The same applies to a loan-application log, where for instance, \emph{finalize loan application} must naturally not precede \emph{create application}.
Conversely, we find that logs for which our framework achieved poor performance correspond to rather specialized processes, for instance, for a log of a process that handles the calculation of compensation during parental leave, our framework only achieved a precision of 0.30 and recall of 0.56 
and for a log that captures the interaction between a technician and a pharmacist with a software system, we achieved a precision of 0.20 and recall of 0.78. 
This indicates that activity-level constraints are especially useful to detect violations in processes executed by many organizations in similar ways, i.e., the main use case for reference process models which our framework expects as input.

\mypartwo{Inter-object constraints}
For inter-object constraint, we observe similar trends as for activity-level ones. The best configuration achieves a precision of 0.34 and a recall of 0.82.
For rather streamlined processes, such as a claim-handling process, it correctly detects that a \emph{claim} should occur before a \emph{payment} and that \emph{payment} and \emph{claim rejection} cannot co-occur. Accordingly, in a travel booking process, it perfectly identifies that a \emph{plane ticket} should be involved in a process before \emph{luggage} is and a \emph{destination} should occur before there is a \emph{plane ticket}. Larger numbers of false positives are detected, for instance, when the process contains many different object types and at the same time there is a lot of variability in the process, i.e., many execution variants. For instance, in a log capturing a heath-care process that spans multiple departments, 349 false positives violations were found, compared to 75 true positive ones.

\mypartwo{Intra-object constraints}
When considering intra-object constraints, we observe substantially better performance compared to the other constraint types. The best configuration achieves a precision of 0.72 and a recall of 0.79. 
For more than 400 logs a perfect recall and a precision above 0.75 was obtained at the same time. The reason for the better results compared to other constraint types can also be observed by inspecting results of individual logs. For instance, we achieve perfect recall and precision for a sales process, for which, e.g., violations were found where an order was archived before it was rejected, an order was both rejected and confirmed, and an order was confirmed before it was received. Similarly, in a credit-application log we found credits that are offered before they are checked and applications that are never assessed. 
These examples show that our framework is capable of identifying violations of general behavioral
relations that make sense, such as that an object should be created before it is checked. 

However, there are also cases for which our framework fails for intra-object violation detection, i.e., misses true violations. For instance, in an event log that captures the process of candidate selection for scholarships. In the process, candidates are first classified and later selected. Our framework failed to recognize violations where one of these were missing or their order was swapped causing false negatives. A possible reason for this may be the similar meaning of the actions \emph{select} and \emph{classify}, which both describe the separation of elements of a group~\cite{malone2003organizing}. Nevertheless, we found that intra-object constraints provide a reliable means to detect best-practice violations in the context of the same type of object.

\mypartwo{Role constraints}
For role constraints, the best configuration achieves a precision of 0.49 and recall of 0.8. 
We also inspected logs for which role constraints performed perfectly. 
For instance, in a credit-check process all violations of wrong roles performing a task were correctly identified, e.g., that calling a client should be performed by \emph{customer contact}, whereas a credibility check should be performed by \emph{risk management} and a payment should be issued by \emph{finance}. 
However, we also observed false positives.
For instance, for a log capturing the interaction of a pharmacist with an information system that performs some tasks autonomously, such as sending and processing notifications, the framework selected constraints that did not correctly identify who (i.e., the pharmacist or the system) should perform which steps of the process.

\mypartwo{Main insights}
Overall, we observe that with a relatively small number of constraints, picked from a large collection of thousands of constraints per type, we detect a substantial amount of true violations, achieving good recall scores across constraint types. Yet, there are also false positives, which in turn lead to lower precision scores. 
For the goal of detecting best-practice violations allowing for false positives while minimizing false negatives can be acceptable, though, given that a user of our approach can easily filter out constraints that lead to false positives by selecting constraint components to avoid when inspecting the detection results.
The results must also be seen in light of the fact that the underlying model collection varies in terms of quality (despite the automated quality checks we performed), abstraction-level, and modeling style. 
The insights from the individual logs suggest that our framework works particularly well for processes that are commonly executed in similar ways in organizations, which is exactly what we target by mining constraints from reference process models. Expectedly, for more specialized and niche processes the performance decreases.

\mypar{Baseline comparison}
\autoref{table:blcomparison} shows the results of the comparison against the baseline. The comparison results presented here focus on intra-object constraints, which the scope of the baseline is limited to. 
The baseline achieves an average precision of 0.80 and recall of 0.20. Our framework surpasses  the baseline for all configurations in terms of recall: When choosing the configuration that yields the best trade-off between precision and recall ($k=100, \tau=0.5$), we achieve a slightly lower precision than the baseline (0.72), but a recall of 0.79, which is almost four times higher. Choosing a configuration with fewer constraints ($k=10, \tau=0.8$), we also surpass the baseline in terms of precision (0.86), while the recall we achieve with that configuration is still three times higher than the baseline's recall.

Our framework's improved performance, especially in terms of recall, can be attributed to its utilization of a broader range of constraints, resulting in an increase in true positives and a decrease in false negatives.
In particular, the baseline focuses on detecting violations that depend on pairwise relations between actions that are applied to the same object, e.g. that \emph{reject} and \emph{accept} exclude each other. The declarative constraint templates we use as a basis can also detect violations independent of such pairwise relations. For instance, we can detect that an order was never created independent of what happens to the order in the remainder of a trace, whereas the baseline can only detect this based on other actions applied to the order, e.g., if an \emph{archive} action is applied to the order and it observed a relation \emph{create} is followed by \emph{archive}. Nevertheless, despite these improvements compared to the baseline, our framework detects a substantial number of false positives, which needs to be addressed in future work. 
Still, for the goal of detecting best-practice violations, our framework is generally better suited than the baseline. This is because we do not want users to miss potentially relevant violations; thus, it makes sense to sacrifice precision for recall. Only when we do this, we achieve a somewhat lower precision than the baseline, for the benefit of a much higher recall.

\begin{table}[!h]
    \centering
    \caption{Results of the baseline comparison only considering intra-object constraints (averaged across logs)}
    \label{table:blcomparison}
    \small
    \begin{tabular}{lrrrrccc}
        \toprule
        \textbf{Approach}  & \textbf{TP} & \textbf{FP} & \textbf{FN} & \textbf{Precision} & \textbf{Recall} \\
        \midrule
        \textbf{Ours} (best trade-off)   & 68.64 & 131.71 & 22.58 & 0.72 & 0.79 \\
        \hphantom{\textbf{Ours}} (best precision)  & 51.69 & 26.46 & 39.53 & 0.86 & 0.59 \\
                        \midrule
        \textbf{Baseline~\cite{VanDerAa2021}} & 35.56 & 26.16 & 55.66 & 0.80 & 0.20 \\
        \bottomrule
    \end{tabular}
\end{table}

\mypar{Run time} We ran our experiments single-threaded on a laptop with a 2 GHz Intel Core i5 processor and 16GB of memory. In this environment, the constraint mining stage, which has to be run only once for a model collection, took 112.5 minutes (99\% of the time was used for constraint refinement).
The average time to run the selection and checking stages for a log was 72.9 seconds, varying between 19.4 and 167.6 seconds depending on the complexity of the input event log, i.e., the number of activities and object types.

\section{Real-life Application}
\label{sec:reallife}
Finally, we applied our framework to real-life event data of a purchasing process and a sales process to highlight its usefulness in practical settings: 
\begin{compactenum}
    \item \emph{Purchasing process}. We use the BPI 2019 challenge~\cite{bpi19}, which captures event data on a purchasing process at a multinational company. We selected those traces from the original log that are supposed to adhere to a 3-ways-match procedure where the invoice is to be recorded after goods receipt. The resulting log contains 15,182 traces, 319,233 events, and 38 event classes.
    \item \emph{Sales process}. We use a proprietary event log of a sales process. The process instances were executed as part of a business-to-business software sales process of a medium-sized European company. In this process, a lead (customer contact) is generated and eventually converted into an opportunity (lead with specific purchase scenario) that is then either closed or turned into a paying customer. This event log cannot be shared for privacy-related ethics and compliance reasons.
\end{compactenum}
For the real-life application scenarios we employed a proprietary reference model collection of ca.\ 4,000 BPMN models that provide vendor-specific best practices.

\mypar{Results}
We discuss the results obtained for the purchasing process and the sales process separately.

\mypartwo{Purchasing process}
Although there is no gold standard available that indicates best-practice violations in this process, our framework was able to detect a range of potentially undesired behaviors, as shown in \autoref{table:realliferesults1}.

\begin{table}[!htbp]
    \centering
        \small
    \begin{tabularx}{\linewidth}{cXXr}
        \toprule
        \textbf{ID} & \textbf{Constraint} & \textbf{Constraint explanation} & \textbf{\#Violations} \\
        \midrule
           \textit{p1} & \textsc{Response}(\emph{create}, \emph{confirm})$\mid$\emph{invoice} & After an \emph{invoice} is \emph{created}, it should be \emph{confirmed}.& 10,722 \\
        \textit{p2} & \textsc{Succession}(\emph{goods receipt}, \emph{invoice}) & An \emph{invoice} occurs iff it is followed by a \emph{goods receipt}.& 9,601	\\
        \textit{p3} & \textsc{AtLeastOne}(\emph{create})$\mid$\emph{invoice} &  For an \emph{invoice}, \emph{create} should occur. & 940 \\
         \textit{p4} & \textsc{Response}(\emph{purchase order item}, \emph{goods receipt}) & A \emph{goods receipt} should follow a \emph{purchase order item}.& 638 \\
         \bottomrule
    \end{tabularx}
    \caption{Exemplary best-practice constraints that were selected by our framework and the number of traces that violated them in the real-life purchasing event log.}
    \label{table:realliferesults1}
\end{table}

The examples correspond to situations where an invoice is never created (p3), where purchase order items are never followed by goods receipt (p4), and invoices that are created but never confirmed (p1). All of these issues represent behavior that deviates from best practices extracted from a high-quality reference model collection that includes models for purchasing, invoicing, and payment processes.  
Most interestingly, in a considerable amount of traces a good receipt is not followed by an invoice (p2), which violates the underlying matching procedure, where invoices should be recorded after goods are received. These may include incomplete traces, yet, may also hint at undesired behavior, which is worth investigating further.

\mypartwo{Sales process}
For the sales process event log we also found some best-practice violations based on constraints recommended by our framework as shown in \autoref{table:realliferesults2}. In particular, there are traces where a lead is never created (s4) and assigned more than once or never (s1). Furthermore, there is a considerable amount of opportunities that are never closed (s2) and activities in the process that are recorded but not documented properly afterwards (s3). 
A discussion with a technical process operations specialist confirmed that all of these violations indicate lacking process discipline in the affected traces, i.e., these do not adhere to established procedures, which may cause efficiency and quality issues in the process. The violations, thus, provide valuable insights into process improvement opportunities. 

\begin{table}[!htbp]
    \centering    
    \small
    \begin{tabularx}{\linewidth}{cXXr}
        \toprule
        \textbf{ID} & \textbf{Constraint} & \textbf{Constraint explanation} & \textbf{\#Violations} \\
        \midrule
        \textit{s1} & \textsc{ExactlyOne}(\emph{assign})$\mid$\emph{lead} &  For a \emph{lead}, \emph{assign} should occur once. &  220,057\\
        \textit{s2} & \textsc{ExactlyOne}(\emph{close})$\mid$\emph{opportunity} &  For an \emph{opportunity}, \emph{close} should occur once.&  11,175\\
         \textit{s3} & \textsc{AlternateResponse}(\emph{activity logged task}, \emph{enter notes})&  \emph{Activity logged task} should be followed by \emph{enter notes}. &  9,228\\
         \textit{s4} & \textsc{AtLeastOne}(\emph{create})$\mid$\emph{lead} &  For a \emph{lead}, \emph{create} should occur. & 418 \\
         \bottomrule
    \end{tabularx}
    \caption{Exemplary best-practice constraints that were selected by our framework and the number of traces that violated them in the real-life sales process event log.}
    \label{table:realliferesults2}
\end{table}

These insights 
further highlight the benefits of our framework to detect potentially undesired behavior that violates best-practices without the need for dedicated process models to be available.

\section{Related Work}
\label{sec:related}
In this section, we discuss the primary research streams to which our work relates: conformance checking,  declarative process mining, and process matching.

\mypar{Conformance checking}
Conformance checking is one of the main tasks of process mining, aiming to detect deviations between true behavior recorded in event logs and desired behavior captured in a process model
\cite{carmona2018conformance, bauer2022sampling}.
If the input models capture rules or regulations, conformance-checking techniques can thus be used for process compliance checking, where violations of these shall be detected~\cite{caron2013compliance}.
Most research on conformance checking focuses on scenarios where an event log and a dedicated imperative process model, e.g., a BPMN diagram, are available. 
In this context, conformance is commonly checked based on so-called alignments~\cite{adriansyah2011conformance, van2012replaying}, where observed  traces are compared with executions allowed by a predefined process model to find deviations. In the context of declarative models, conformance checking boils down to checking for each trace in an event log, if it satisfies each constraint in a declarative process model~\cite{DiCiccio2022, burattin2016conformance}. 
To get a global view on conformance, the idea of aligning log traces to the closest \emph{model trace} (considering all model constraints) can also be lifted to a declarative setting~\cite{deleoni2015declaratvealignments}. 
In our framework, we adopt the former approach, though, because we do not check conformance against a given normative (declarative) model. Instead, we mitigate the need for such a dedicated model and rather check against a collection of relevant, yet individual, best-practice constraints, essentially querying the event log (cf.~\cite{polyvyanyy2017querying}) for violating traces.

\mypar{Declarative process mining}
Declarative process mining is an active research field and we refer the interested reader to a recent overview of the topic~\cite{DiCiccio2022}. 
In this context, \textsc{Declare} is among the most prominent formalisms.
It relies on LTL, leveraging its reasoning mechanisms for, among others, model analysis and conformance checking, which we utilize in our framework.
Research in this area is also dedicated to identifying and resolving inconsistencies and redundancies in declarative process models \cite{diciccio2017, corea2022measuring, corea2021interactive}, which we leverage for both refining our constraint collection in the \emph{Constraint Mining} stage of the framework and for checking if the set of recommended constraints is actually consistent in the \emph{Constraint Selection} stage.

Recently, \emph{object-centric behavioral constraints} have been proposed, which allow for a detailed specification of complex networks of objects and their relations that co-evolve in one or multiple processes~\cite{artale2019enriching}. 
While still in its infancy~\cite{DiCiccio2022}, this line of research is promising to be adopted for our framework. The level of detail this formalism can express goes beyond what is typically captured in BPMN diagrams, though.
Therefore, we would first need to extend the scope of the input of our framework, which we aim to address in future work.

Research on declarative process mining further focuses on discovering declarative models from event logs~\cite{maggi2012efficient, ciccio2015discovery}. In this context, measuring the interestingness of traces of an event log given a declarative process model~\cite{cecconi2018interestingness} and vice versa~\cite{cecconi2022measuring} is researched. This closely relates to the notion of relevance we employ in our framework, where we essentially want to assess how interesting a constraint is for a given log. Complementary to existing approaches, we consider semantic similarity as an indicator of interestingness. 

\mypar{Process matching}
Process matching techniques revolve around establishing connections between process concepts found in various artifacts. 
The primary research focus lies in process model matching, where the objective is to create links between activities present in different process models. 
To achieve this, process model matchers leverage diverse process model features including model structure~\cite{dijkman2011similarity} and allowed behavior~\cite{kunze2011behavioral}, but also natural-language-based features~\cite{ehrig2007measuring}. 
Beyond model-to-model matching, there is also  work on matching (parts of) models to events recorded in event logs~\cite{baier2014bridging, van2017instance}.
The selection stage of our framework also creates links between components of the constraints and event log counterparts based on semantic similarity in order to select the most relevant constraints for the given log.

\section{Conclusion}
\label{sec:conclusion}

\mypar{Summary}
In this paper, we propose a framework to mine declarative constraints from reference process models to detect best-practice violations in event logs. 
Our framework extracts and refines constraints based on imperative models, instantiates and selects relevant constraints given an event log, and checks whether the process executions recorded in the log violate the best-practices captured by the constraints. 
In this manner, we mitigate the need for process models specifically designed to capture the desired behavior of the process recorded in the given event log.
Our experiments show that with a small number of constrains, which our framework selects from a collection of tens of thousands of mined constraints, we correctly detect a substantial amount of violations. Furthermore, application scenarios based on real-life reference models and event logs demonstrate the framework's usefulness in practical settings.

\mypar{Limitations}
Our framework and evaluation are subject to certain limitations. 
First, our framework does not cover constraints for all process perspectives, excluding constraints related to the time and data perspectives from consideration, 
 e.g., that an order should be approved within a given amount of time or that a specific check should be performed for orders of a certain value. 
This is because such constraints are rarely found in reference process models, whereas they are also much harder to generalize than our activity-, object-, and role-related ones, because restrictions on time and monetary values can differ substantially across organizations and processes. 
Second, in order to provide relevant best-practice violations, the event log is assumed to be on the same granularity level as the components of constraints the framework mined. If the event log records activities on a more fine-granular level, it may, therefore, be necessary to first abstract the event log before applying our framework.
With respect to our evaluation, there are certain threats to validity. 
A threat to internal validity is that the process models used as a basis for our experiments, although being real-life models, stem from different sources and were established for different purposes, which means that they jointly do not represent a \textit{reference} model collection (due to the absence of publicly available reference collections).
Despite automated quality checks, the contained models may thus vary in terms of quality, granularity, and modeling style. A threat to external validity is the (currently) limited application of our framework in practice. While we applied it on real-life event data, which yielded interesting best-practice violations that were verified with a specialist, it has not been field-tested in an organization so far.

\mypar{Future work}
In future work, we aim to address the aforementioned limitations of the framework and its evaluation, as well as provide further extensions. This primarily includes the following.
First, we aim to broaden the scope of input beyond imperative process models, for instance by leveraging data models like entity-relationship diagrams and database schemata for constraint mining, since these can give complementary insights into relationships between individual objects.
On the output side, we aim to extend the framework to provide insights about potential root causes of violations by connecting violated constraints with shared parameters to each other. For instance, given two violated constraints, where one captures that an invoice should be checked at least once and the other one that an invoice should be checked before it is approved, we aim to derive that a missing invoice check is the cause of both violations.
Finally, we aim to implement the framework into a tool including a graphical user interface. This will allow users to adapt and specify their own constraints according to their needs, based on event data and mined constraints, as well as to analyze constraints and their violations in a more interactive manner. Based on this tool, we aim to conduct a user study asking users to apply the framework in an organizational setting using their own event logs in order to further assess the practical value of our work.



\bibliographystyle{elsarticle-num}
\bibliography{refs}

\end{document}